\documentclass{aa}  

\usepackage{savesym}
\usepackage{amsmath}
\savesymbol{iint}
\usepackage{txfonts}
\restoresymbol{TXF}{iint}
\usepackage{graphics,graphicx}
\usepackage{amssymb}
\usepackage{color}
\usepackage[draft]{hyperref}
\usepackage{breakurl}
\usepackage{array}
\usepackage{rotating}
\usepackage{xcolor}
\usepackage{natbib}
\usepackage{float}
\usepackage{flafter}
\usepackage{booktabs}
\usepackage{afterpage}
\usepackage{lipsum}
\usepackage{longtable}
\usepackage{lscape}
\bibpunct{(}{)}{;}{a}{}{,}

\newcommand{\gal}{\object{M\,31}}

\newcommand{\chandra}{{\it Chandra}}
\newcommand{\xmm}{{\it XMM-Newton}}
\newcommand{\rosat}{{\it ROSAT}}
\newcommand{\einstein}{{\it Einstein}}

\newcommand{\spitzer}{{\it Spitzer}}
\newcommand{\galex}{{\it GALEX}}
\newcommand{\herschel}{{\it Herschel}}
\newcommand{\hst}{{\it HST}}

\newcommand{\phabs}{{\tt phabs}}

\newcommand{\apec}{{\tt apec}}
\newcommand{\nei}{{\tt nei}}

\newcommand{\nh}{{$N_{\mathrm{H}}$}}

\newcommand{\ha}{H$\alpha$}
\newcommand{\hi}{$\ion{H}{i}$}
\newcommand{\hii}{$\ion{H}{ii}$}
\newcommand{\sii}{[$\ion{S}{ii}$]}

\newcommand{\oiii}{[$\ion{O}{iii}$]}

\begin{document} 

   \title{Deep XMM-Newton observations of the northern disk of M31 II: Tracing the hot interstellar medium\thanks{Based on observations obtained with \xmm, an ESA science mission with instruments and contributions directly funded by ESA Member States and NASA}}

          
   \author{ Patrick~J.~Kavanagh\inst{1} 
            \and 
            Manami~Sasaki\inst{2} 
            \and
            Dieter~Breitschwerdt\inst{3}
            \and
            Miguel~A.~de~Avillez\inst{4}
            \and
            Miroslav~D.~Filipovi\'c\inst{5}
            \and
            Timothy~Galvin\inst{5}
            \and
            Frank~Haberl\inst{6}
            \and
            Despina~Hatzidimitriou \inst{7,8}
            \and
            Martin~Henze\inst{9}
            \and
            Paul~P.~Plucinsky\inst{10}
            \and
            Sara~Saeedi\inst{11}
            \and
            Kirill~V.~Sokolovsky\inst{12,13,14}
            \and
            Benjamin~F.~Williams\inst{15}
          }


\institute{
School of Cosmic Physics, Dublin Institute for Advanced Studies, 31 Fitzwillam Place, Dublin 2, Ireland, \email{pkavanagh@cp.dias.ie}
\and 
Remeis Observatory and ECAP, Universit\"{a}t Erlangen-N\"{u}rnberg, Sternwartstra{\ss}e 7, D-96049 Bamberg, Germany
\and
Zentrum f\"{u}r Astronomie und Astrophysik, Technische Universit\"{a}t Berlin, Hardenbergstra{\ss}e 36, D-10623 Berlin, Germany
\and 
Department of Mathematics, University of Évora, R. Romão Ramalho 59, 7000 Évora, Portugal
\and
Western Sydney University, Locked Bag 1797, Penrith South DC, NSW 1797, Australia
\and
Max-Planck-Institut f\"{u}r extraterrestrische Physik, Gie{\ss}enbachstra{\ss}e 1, D-85748 Garching, Germany
\and
IAASARS, National Observatory of Athens, 15236 Penteli, Greece
\and
Department of Astrophysics, Astronomy \& Mechanics, Faculty of Physics, University of Athens, 15783 Athens, Greece
\and
Institute of Space Sciences (IEEC-CSIC), Campus UAB, Carrer de Can Magrans, s/n 08193 Barcelona, Spain
\and
Harvard-Smithsonian Center for Astrophysics, 60 Garden Street, Cambridge, MA 02138, USA
\and
Institut f\"ur Astronomie und Astrophysik, Universit\"at T\"ubingen, Sand 1, D-72076 T\"ubingen, Germany
\and
Department of Physics and Astronomy, Michigan State University, 567 Wilson Rd, East Lansing, MI 48824, USA
\and
Sternberg Astronomical Institute, Moscow State University, Universitetskii pr. 13, 119992 Moscow, Russia
\and
Astro Space Center of Lebedev Physical Institute, Profsoyuznaya St. 84/32, 117997 Moscow, Russia
\and
Astronomy Department, University of Washington, Box 351580, Seattle, WA 98195, USA
}

\date{Received ??; accepted ??}

\abstract{}{We use new deep \xmm\ observations of the northern disk of \gal\ to trace the hot interstellar medium (ISM) in unprecedented detail and to characterise the physical properties of the X-ray emitting plasmas.}{We used all \xmm\ data up to and including our new observations to produce the most detailed image yet of the hot ISM plasma in a grand design spiral galaxy such as our own. We compared the X-ray morphology to multi-wavelength studies in the literature to set it in the context of the multi-phase ISM. We performed spectral analyses on the extended emission using our new observations as they offer sufficient depth and count statistics to constrain the plasma properties. Data from the Panchromatic Hubble Andromeda Treasury were used to estimate the energy injected by massive stars and their supernovae. We compared these results to the hot gas properties.}{The brightest emission regions were found to be correlated with populations of massive stars, notably in the 10~kpc star-forming ring. The plasma temperatures in the ring regions are $\sim0.2$~keV up to $\sim0.6$~keV. We suggest this emission is hot ISM heated in massive stellar clusters and superbubbles. We derived X-ray luminosities, densities, and pressures for the gas in each region. We also found large extended emission filling low density gaps in the dust morphology of the northern disk, notably between the 5~kpc and 10~kpc star-forming rings. We propose that the hot gas was heated and expelled into the gaps by the populations of massive stars in the rings.}{It is clear that the massive stellar populations are responsible for heating the ISM to X-ray emitting temperatures, filling their surroundings, and possibly driving the hot gas into the low density regions. Overall, the morphology and spectra of the hot gas in the northern disk of \gal\ is similar to other galaxy disks.}

\titlerunning{Tracing the hot ISM of \gal}
\maketitle 

\section{Introduction}
The evolution of spiral galaxies is driven by star formation and the matter cycle between the stars and the interstellar medium (ISM). Massive OB stars inject energy into the ISM through their radiation, stellar winds, and, finally, by supernova (SN) explosions. These processes are often correlated in space and time, producing superbubbles with sizes of typically 100 -- 1000 pc. However, as we are located inside the Galactic disk, we are not able to observe the ISM in the Milky Way with all these processes and resulting structures in their entirety. Instead, for a detailed study of the ISM, we have to look beyond our Galaxy.

The Andromeda galaxy (\gal) is the largest galaxy in the Local Group and the nearest spiral galaxy to the Milky Way, located at a distance of 783~kpc \citep{Conn2016}. With a size and a mass comparable to that of our Galaxy, this archetypal spiral galaxy provides a unique opportunity to study and understand the nature and the evolution of a galaxy similar to our own. Based on \spitzer\ data, \citet{Gordon2006} showed that \gal\ has spiral-arm structures merged with the prominent star-forming ring at a radius of $\sim10$~kpc. The \herschel\ images show a radial gradient in the gas-to-dust ratio and indicate that there are two distinct regions in \gal\ with different dust properties inside and outside a radius $R \approx 3.1$~kpc \citep{Smith2012, Fritz2012}. \citet{Galvin2012} and \citet{Galvin2014} presented high resolution radio continuum images of \gal\ from the Very Large Array, identifying 916 unique discrete radio sources across the field of \gal\ and faint extended emission at the location of the 10~kpc ring.

The star-formation history (SFH) in \gal\ has been studied in detail in observations with both the {\it Hubble Space Telescope} \citep[\hst,][]{Lewis2015} and large ground-based telescopes \citep[e.g. the Local Group Galaxy Survey, LGGS,][]{Massey2006}. \citet{Williams2003} measured a mean star-formation rate of about 1~M$_{\sun}$~yr$^{-1}$ in the full disk of \gal\ over the last 60~Myr and produced maps of star-formation rate in different age ranges. Furthermore, the northern disk was observed in the Panchromatic Hubble Andromeda Treasury (PHAT) survey \citep{Dalcanton2012} resolving over 100~million stars in \gal\ in the near-infrared (NIR) to the ultraviolet (UV).  Using these data, \citet{Lewis2015} showed that the SFH varies significantly on small spatial scales in the northern disk. They found that the 10~kpc ring is at least 400~Myr old with ongoing star formation and estimate about 60\% of all star formation in the last 500~Myr occuring within the ring. By extrapolating the results from the PHAT survey area, the total star formation rate in the disk was estimated to be $\sim0.7$~M$_{\sun}$~yr$^{-1}$ over the last 100~Myr, consistent with previous broadband estimates.

In X-rays, a survey of the entire \gal\ was performed with \xmm\ \citep{Stiele2011}. These \xmm\ observations have revealed an extended diffuse emission in \gal, in particular on the northern side of the galaxy, which is well correlated with the star-forming ring. 
We carried out additional deep observations in two fields of the northern disk to perform a detailed study of the X-ray source population and the diffuse emission. 
Based on the new \xmm\ data, we created a 
catalogue of X-ray sources down to luminosities of $\sim$7$\times$10$^{34}$ erg s$^{-1}$ (0.5 -- 2.0 keV) and improved the classification of the previously known X-ray sources. 
The results of the the point sources study (catalogue, spectral and timing studies, multi-wavelength cross-correlation, etc.), based on the new \xmm\ observations, have been published by \citet[][hereafter Paper~I]{2018A&A...620A..28S}, 
in combination with the results of surveys performed with the \chandra\ X-ray Observatory and the PHAT data taken with the \hst\ \citep{2018ApJS..239...13W}.
In this paper, we present an analysis of these new deep \xmm\ pointings ($\sim200$~ks on each field) 
focusing on 
the diffuse emission 
in order to 
study the structure and the physical properties of the hot ISM plasma. 
The interstellar space in galaxies like the \gal\ and the Milky Way are filled with cool clouds ($T \lesssim 10^{2}$ K) of neutral hydrogen embedded in warm ($T \approx 10^{4}$ K) intercloud medium of partially ionised hydrogen \citep[][and references therein]{Cox2005}. 
At the distance of \gal, $1\arcsec$ corresponds to 3.8~pc, so the spatial resolution of \xmm\ allows us to resolve the hot gas in bubbles and superbubbles in \gal\ on 100 -- 1000~pc scales and study different morphologies, which might be indicative of inhomogeneities in the ISM. The spectral resolution of the EPIC instruments allows us to investigate variations in the physical properties of the ISM such as luminosity, temperature, and metallicity, as well as absorption by cold ISM in the foreground. 
The distribution and the filling factor of the hot phase of the ISM can be studied in detail by comparing the X-ray emission to \hi\ \citep[e.g.][]{Braun2009}, \ha\ (LGGS) data, and cold gas and dust \citep{Fritz2012}. 

We present the \xmm\ data used for this work and the analysis methods applied to the study of the diffuse X-ray emission in Sect.\,2. Also, we discuss the contribution of the unresolved point sources to the observed diffuse X-ray emission. In Sect.~3, we give an overview of the additional multi-wavelength data. In Sect.\,4, morphological studies and spectral analyses are presented, followed by a study of the stellar population in each of the regions and the energy budget in Sect.\,5. A summary and conclusions are presented in Sect.\,6.

\section{Observations and data reduction}
\subsection{X-ray}
We performed deep observations of two fields in the northern disk of \gal\ with \xmm\ \citep{Jansen2001} as part of our Large Project survey (PI M. Sasaki), details of which can be found in Paper~I (see also Table~\ref{obscat}). In the following sections we describe the reduction of these data and supplementary X-ray data for our imaging and spectroscopic analyses.

\subsubsection{Imaging}
In addition to our new deep fields, \xmm\ has observed \gal, including the northern disk, on many occasions over its lifetime. The majority of these observations focus on the core region, but a survey of \gal\ has also been performed covering the ellipse enclosed by the $D_{25}$ isophote down to a limiting luminosity of $10^{35}$~erg~s$^{-1}$ in the 0.2--4.5 keV band \citep{Stiele2011}. In addition, various interesting objects located around \gal\ have also been covered in separate, dedicated observations. In order to properly reveal and characterise the hot ISM plasma in \gal\ we made use of all available \xmm\ data taken up to and including our deep exposures of the northern disk. The area of \xmm\ coverage used for our \gal\ mosaic corresponds to that of \citet{Stiele2011}. This coverage is plotted in relation to the stellar population in their Fig.~1, with regions N1 and N2 roughly corresponding to our new deep fields in the northern disk.

To produce mosaics of \gal\ and its northern disk we used the mosaicing capabilities of the \xmm\ Extended  Source  Analysis  Software (XMM-ESAS), packaged in SAS\footnote{Scienctific Analysis Software, see \burl{http://xmm.esac.esa.int/sas/}}~15.0.0. XMM-ESAS is based on the software used for the background modelling described by \citet{Snowden2004}, and comprises a set of tasks to produce images from observational data and to create model quiescent particle background (QPB) images that can be subtracted from the observational science products \citep[see][]{Kuntz2008,Snowden2008}.
 
We considered only those observations where the primary instrument was the European Photon imaging Cameras (EPICs), which use a pn-type CCD \citep{Struder2001} or a MOS-type CCD \citep{Turner2001} imaging spectrometers (one EPIC with pn CCD and two with MOS CCD). We used only EPIC observations taken in Full Frame mode and processed each of the observational datasets according to the ESAS Cookbook\footnote{See \burl{http://heasarc.gsfc.nasa.gov/docs/xmm/esas/cookbook/xmm-esas.html}}. Standard filtering and calibration were applied using the \texttt{epchain}, \texttt{emchain}, \texttt{pn-filter}, and \texttt{mos-filter} tools. Finally, we discarded EPIC-pn and EPIC-MOS observations with soft proton (SP) flare-filtered exposure times less than 15~ks. Table~\ref{obscat} summarises all the data included in our final screened sample. 

The XMM-ESAS task \texttt{cheese-bands} was used to produce source exclusion masks for each observation. This task performs the  source  detection  on  all  available  EPIC  instruments in multiple energy bands for a given observation and determines source exclusion regions accounting for the shape of the point spread function at the source location on the detector. We chose the 0.3--2.0~keV and 2.0--7.0~keV bands so that the source detection would be sensitive to both soft and hard sources and defined the exclusion limit as 1.1 times the W90 radius (the 90\% enclosed energy radius). Source lists from individual observations were merged using the \texttt{merge\_source\_list} task, and a final source mask was created using the \texttt{make\_mask\_merge} task. Bright sources at or near the edges of individual fields-of-view (FOVs) presented a problem, however.  Uncertainty in source centroids resulted in the \texttt{merge\_source\_list} task not recognising that what appeared to be two separate sources were in fact the same source. In these cases we manually added the exclusion regions to the masks.

\begin{figure*}[!ht]
\begin{center}
\resizebox{\hsize}{!}{\includegraphics[trim= 0cm 0cm 0cm 0cm, clip=true, angle=0]{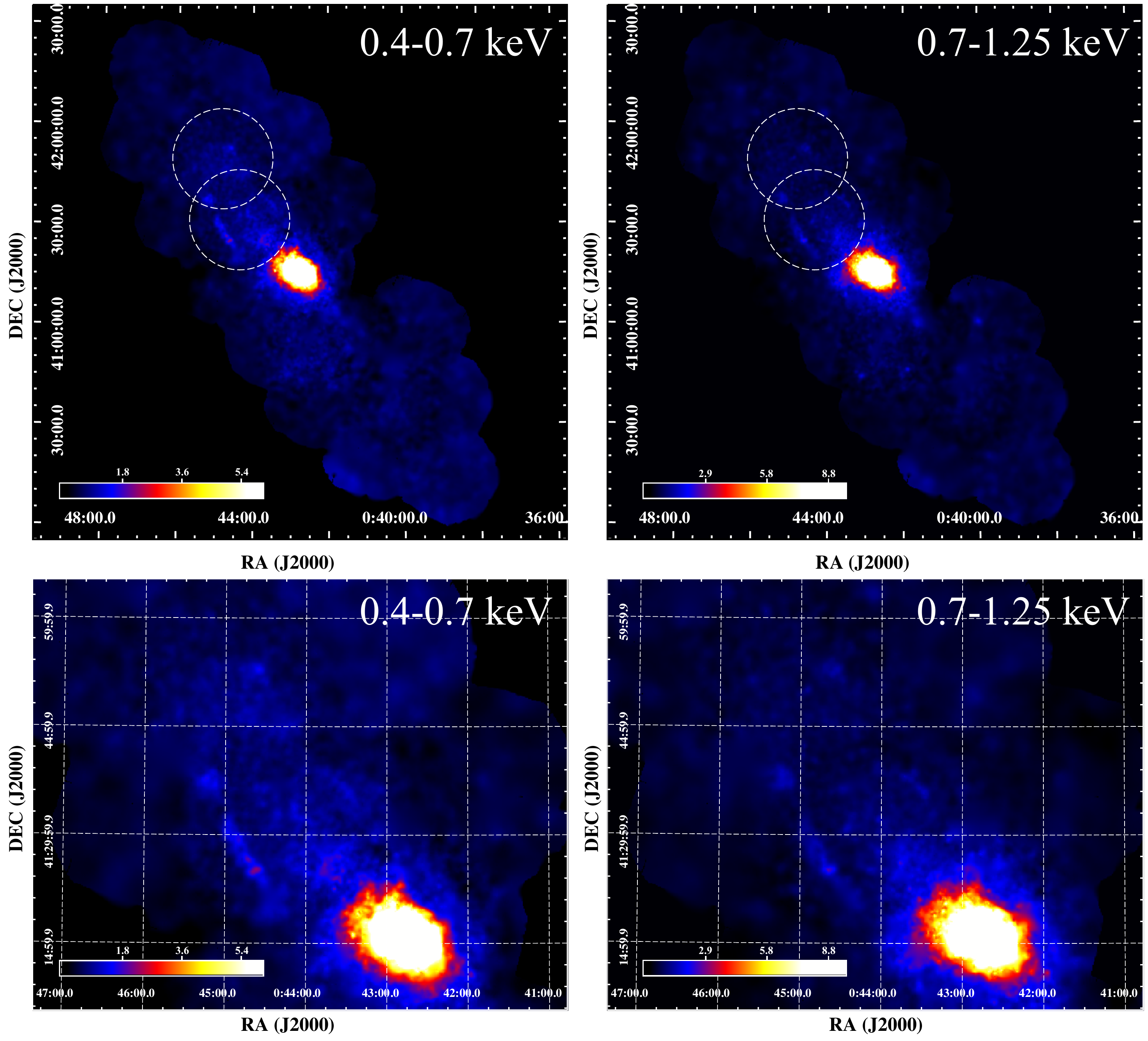}}
\caption{\xmm\ mosaic images of \gal\ with 0.4--0.7~keV shown in top-left and 0.7--1.25~keV in top-right. The white dashed circles mark the fields of our northern disk LP survey, which are revealed in more detail in the zoomed images in bottom-left and bottom-right. The \xmm\ coverage in relation to the stellar population is shown in Fig.~1 of \citet{Stiele2011}. The colours are in linear scale as shown by the colour bars in each image. The units of the images are cts~s$^{-1}$~deg$^{-2}$.}
\label{xmm_images}
\end{center}
\end{figure*}

The \texttt{pn-spectra} and \texttt{mos-spectra} tasks were then used to produce images in the 0.3--0.7~keV, 0.7--1.25~keV, and 2.0--4.5~keV, with point sources masked. We excluded the 1.25--2.0~keV range to remove the contribution of the strong Al~K$\alpha$ and Si~K$\alpha$ instrumental fluorescence lines from the images. The \texttt{pn-back} and \texttt{mos-back}
tasks were then used to produce the corresponding QPB images. Using XMM-ESAS, it is also possible to model and subtract residual soft-proton (SP) contamination from the images. This is important when producing large mosaics as some observations are affected by residual SP while others are not. To this end we extracted full-field spectra and QPB backgrounds for each observation and fitted the spectra in the 3--7~keV energy range where the SP contamination is expected to dominate. In observations where residual SP contamination was identified, the determined spectral parameters were used to generate model SP contamination images with the XMM-ESAS task \texttt{proton}.
Subsequently, we used the \texttt{merge\_comp\_xmm} task to create count, exposure, QPB, and residual SP mosaic images. Finally, the \texttt{adapt\_merge} task was used to create particle background subtracted and exposure corrected mosaics in each energy band, binned into $\sim10\arcsec\times10\arcsec$ bins and adaptively smoothed to yield the final images. To aid in the visualisation of the diffuse emission in \gal, `holes' due to masked point sources were filled using the biharmonic function based inpainting algorithm\footnote{See \burl{http://scikit-image.org/docs/dev/auto_examples/filters/plot_inpaint.html} for description.} contained in the scikit-image python package\footnote{\burl{http://scikit-image.org/}}. These mosaics are presented in Fig.~\ref{xmm_images}.

\subsection{X-ray morphology}
\label{nd_xray}
\begin{figure*}
\begin{center}
\resizebox{\hsize}{!}{\includegraphics[trim= 0.0cm 0.0cm 0.0cm 0.0cm, clip=true, angle=0]{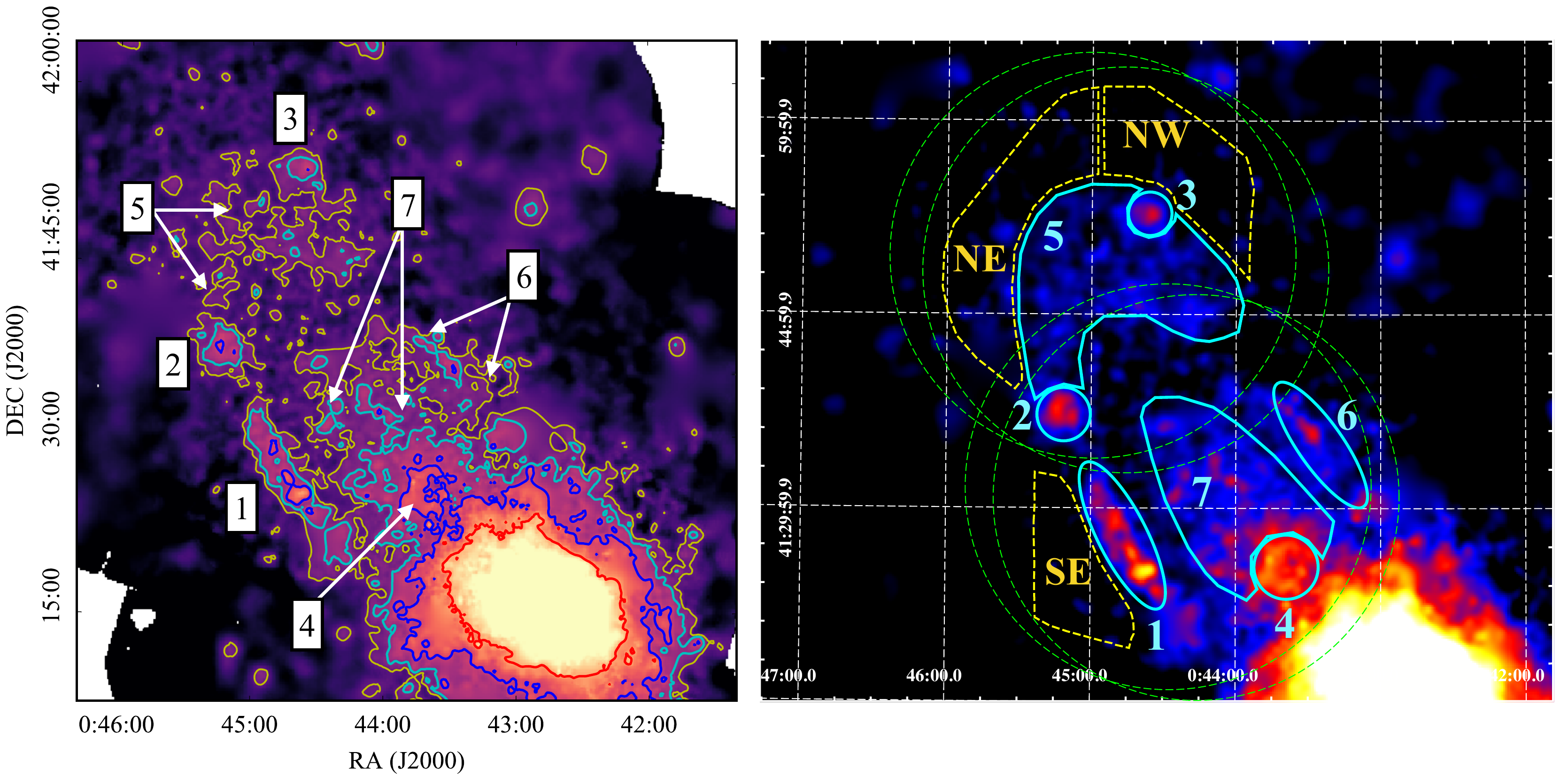}}
\caption{Left: \xmm\ 0.4--1.25~keV image of the northern disk of \gal\ with contours. The image has been adaptively smoothed and masked point sources filled as described in the text. Yellow, cyan, blue, and red contours correspond to $5\sigma$, $10\sigma$, $20\sigma$, and $50\sigma$ above the average background level determined from regions exterior to the northern disk. Regions of interest discussed in the text are marked. Right: \xmm\ 0.4--1.25~keV image of the northern disk with the spectral extraction regions defined from the contours in left shown in cyan. Background regions are shown in yellow. The deep EPIC northern disk fields are shown by the green circles.}
\label{contour_plot}
\end{center}
\end{figure*}

To identify regions of \gal\ with significant X-ray emission, we determined the average background surface brightness in our mosaic and corresponding standard deviation ($\sigma$) in the 0.4--1.25~keV band from several background regions exterior to the northern disk. We then defined contours at $5\sigma$, $10\sigma$, $20\sigma$, and $50\sigma$ above the background level. The resulting image and contours are shown in Fig.~\ref{contour_plot}-left. The X-ray contours follow the spiral structure of \gal\ (see discussion on multi-wavelength morphology in Sect.~\ref{mwm} and Fig.~\ref{multiwave_images}). 

Three bright regions of X-ray emission are evident along the bright star-forming ring which are most likely related to high mass stellar populations (see Sects.~\ref{mwm} and \ref{stellar_input} for further details). We labelled these as Regions 1-3 in Fig.~\ref{contour_plot}-left. A fourth bright emission region is seen closer to the core of \gal\, though still appearing to be associated with a massive stellar population. We labelled this as Region 4 (Fig.~\ref{contour_plot}-left). Larger, fainter extended emission is observed just inside the star forming ring, filling the gap between the ring and inner dust lane (see Fig.~\ref{multiwave_images}). This emission is broken towards the north. Therefore, we separated this ring-adjacent emission into two called Regions 5 and 6, as shown in Fig.~\ref{contour_plot}-left. The faint extended emission inside the dust lanes was labelled Region 7. We used the 5~$\sigma$ contours to define spectral extraction regions for these extended emission regions, which are shown in Fig.~\ref{contour_plot}-right.

The brightest emission in our mosaic was found in the nuclear region. While some of this emission is likely diffuse in origin \citep{Li2007}, much of the emission is due to the unresolved population of X-ray emitting objects such as cataclysmic variables (CVs), active binaries (ABs), X-ray binaries (XRBs), etc. In addition, the multitude of very bright and variable X-ray sources in and around the nucleus, and the relatively large \xmm\ point-spread-function, makes the spectral extraction and analysis of this region very challenging. For this reason, we do not analyse the emission from the \gal\ core.

\subsubsection{Spectra}
\label{spec_analysis}
The mosaics constructed from the available data showed diffuse structure in the northern disk of \gal\ only. This is due to our new deep observations of this region and highlights the necessity of similarly deep observations in the southern disk to reveal its hot ISM plasma. Therefore, our spectral analysis is focused solely on the northern disk viewed through our observations. We note that the spectral analysis presented in this work is somewhat complex with multiple thermal and non-thermal components contributing to a single spectrum with these having instrumental, particle, and astrophysical origins. Therefore, particularly given the faint nature of the emission being studied, it is essential to constrain the various contributions to our spectra as much as possible. In this section we outline our approach to extracting and analysing the spectra, and constraining the instrumental, astrophysical background, and unresolved \gal\ source contributions.

For the spectral extraction we closely followed the procedures discussed in detail in
\citet{Maggi2016}, only differing in our approach to the background fitting. In this work, we first constrained the background emission, then fixed these emission components in fits of the extended emission region spectra, rather than fit spectra of extended and background regions simultaneously. We used standard tools in SAS to produce our spectra and associated files. Prior to extracting spectra, we generated vignetting-weighted event lists for the EPIC instruments to correct for the effective area variation across the FOV using the SAS task \texttt{evigweight}. Spectra were extracted from our extended emission regions (see Sect.~\ref{nd_xray}), with background spectra extracted from regions within our deep fields (see Sect.~\ref{x-ray_background}). Out-of-time spectra were also created, scaled, and subtracted from the extended and background spectra following standard SAS procedures\footnote{\burl{https://www.cosmos.esa.int/web/xmm-newton/sas-thread-epic-oot}}. The point source catalogue from Paper~I was used as input to the SAS task \texttt{region} to determine source exclusion regions. Exclusion regions were defined using the 90\% encircled energy fraction of the sources. Extended and background spectra were extracted using the SAS task \texttt{evselect}, with redistribution matrices and ancillary response files produced by the SAS tasks \texttt{rmfgen} and \texttt{arfgen}, respectively. The spectra were grouped to minimum of 30 counts per bin to allow the use of the $\chi^{2}$-statistic. All fits were performed using XSPEC \citep{Arnaud1996} version 12.9.1 with ATOMDB version 3.0.3\footnote{\burl{http://www.atomdb.org/}}. Abundance tables were set to those of \citet{Wilms2000} and photoelectric absorption cross-sections to those of \citet{Bal1992}. 

\subsubsection{X-ray background}
\label{x-ray_background}
We were somewhat limited in our choice of background regions in the northern disk fields given that our extended emission regions cover most of the EPIC instrument FOVs. We were further constrained as we required regions that fell on the detectors in both deep pointings at each northern disk field (see Fig.~\ref{contour_plot}-right green circles). This is important for identifying any time variable background components, such as residual SP contamination and Solar Wind Charge Exchange (SWCX) emission. Three background regions were chosen which we hereafter refer to as SE, NE, and NW, and are shown in Fig.~\ref{contour_plot}-right. Spectral fits were performed on these regions to characterise the X-ray background components.

\paragraph{Astrophysical components:}
The astrophysical X-ray photon background (AXB) typically comprises four or fewer components \citep{Snowden2008,Kuntz2010}, namely the unabsorbed thermal emission from the Local Hot Bubble (LHB, $kT \sim 0.1$~keV, fixed in the model), an absorbed cool Galactic halo ($kT_{cool} \sim 0.1$~keV), a higher temperature ($kT_{hot} \sim0.2-0.7$ keV), absorbed thermal component representing emission from the hotter Galactic halo or intergalactic medium (IGM), and an absorbed power law representing unresolved background active galactic nuclei (AGN). To account for the unresolved AGN in our fits, the photon index ($\Gamma$) and normalisation of the power law were fixed to $\Gamma = 1.46$ and the equivalent of 10.5~photons~cm$^{-2}$~s$^{-1}$~sr$^{-1}$ at 1~keV \citep{Chen1997}, respectively. In the case of our observations, halo emission from \gal\ may also supplement the Galactic halo and IGM emission. Rather than explicitly model this, we simply let the two thermal components represent the combined emission from the Galaxy, \gal, and the IGM. All thermal components were fitted with the \apec\ \citep{Smith2001} thermal plasma model in XSPEC. Photoelectric absorption models (the \phabs\ model in XSPEC) were included to account for absorption in both the Galaxy and \gal. The Galactic value was fixed to $1\times10^{21}$~cm$^{-2}$ based  on the \citet{Dickey1990} \hi\ maps, determined using the HEASARC $N_{\rm{H}}$ Tool. We also note that X-ray photons resulting from SWCX can affect the fitting and interpretation of thermal plasma model applied to sources of interest. This time variable emission results from solar wind ions interacting with neutral atoms in the heliosphere or Earth's magnetosheath. This emission is characterised by emission lines from the ion species in the solar wind, such as C, O, Mg, etc., which can masquerade as thermal emission lines. In trial fits to our spectra we searched for any variation between pointings at our northern disk fields but found no evidence of emission line variation. We concluded that SWCX does not significantly affect our spectra and ignored it during subsequent analyses.

\paragraph{Particle induced components:}
The particle-induced background of the EPIC consists of the QPB, instrumental fluorescence lines, electronic read-out noise, and residual SP contamination. To determine the contribution of the first three components we made use of vignetting corrected filter wheel closed (FWC) data. Due to the very strong and position dependent contribution of electronic noise below 0.4~keV and the strong fluorescence emission line contribution above 7~keV for the EPIC-pn detector, we limit our spectral analysis to the 0.4--7~keV range. We extracted spectra from  FWC data from corresponding regions on the detector as our extended regions and the backgrounds. The EPIC-pn and EPIC-MOS FWC spectra were fitted with the empirical models developed by \citet{SturmPhD} and \citet{Maggi2016}, respectively. Since these spectral components are not subject to the instrumental response, we used a diagonal response in XSPEC. The FWC data were fitted simultaneously with the corresponding source and background spectra, with FWC component parameters linked between the two. The only variables were the widths and normalisations of the strongest fluorescence lines, as well as a scaling factor to account for any difference in the QPB level between the FWC and extended and background spectra. Residual SP contamination is a time and position dependent component of the particle background. We allowed for this by including a power law not convolved with the instrumental response in the fits to the source and background spectra \citet{Kuntz2008}.

\begin{figure*}
\begin{center}
\resizebox{\hsize}{!}{\includegraphics[trim= 0cm 0cm 0cm 0cm, clip=true, angle=0]{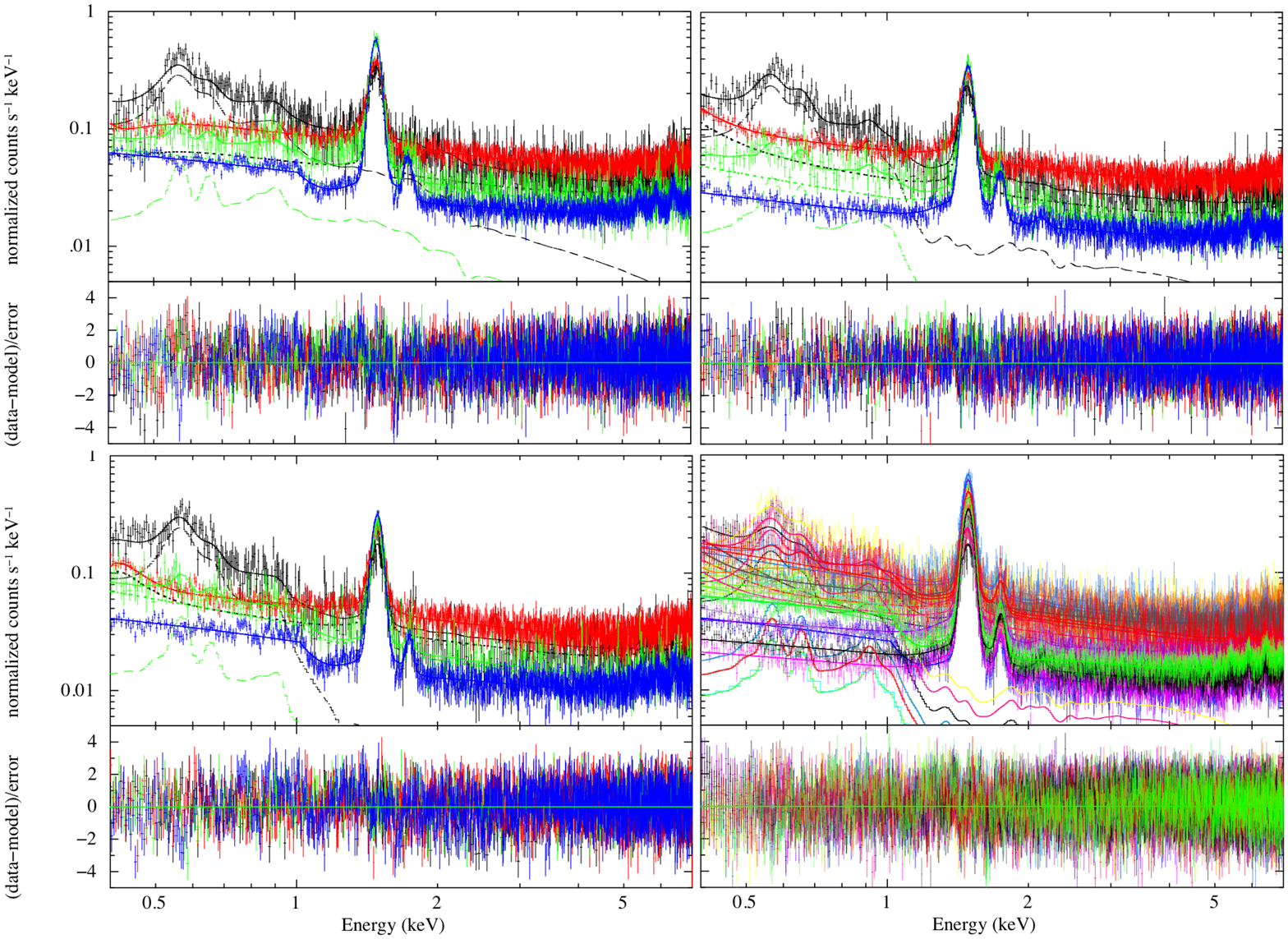}}
\caption{\xmm\ spectral fits of the background regions. \textit{Top-left}: simultaneous spectral fits to the background region NE, and corresponding FWC spectra extracted from EPIC-MOS2 (green and blue) and EPIC-pn (black and red). The dashed lines show the combined AXB contribution while the dotted lines show the combined particle-induced contribution to the extended emission spectra.} For clarity, only spectra from ObsID~0763120301 are shown. \textit{Top-right}: Same as top-left but for background region NW. \textit{Bottom-left}: Same as top-left but for background region SE. In this case the spectra from ObsID~0763120101 are shown. \textit{Bottom-right}: Simultaneous fits to all background spectra with each background region source and FWC spectrum shown in a different colour. The combined AXB and FWC contributions are also shown for each. See text for details.
\label{bkg_spectra}
\end{center}
\end{figure*}

We fitted the background spectra simultaneously with their corresponding FWC spectra to account for the particle-induced counts. Multiplicative factors accounting for the area of the spectral extraction regions in arcmin$^{2}$ were included in the XSPEC model, normalising the astrophysical model components to acrmin$^{2}$. These factors were derived from the \texttt{BACKSCAL} keyword set in the metadata during processing, which takes into account missing sky area due to chip gaps, bad pixels, etc. The fit results are given in Table~\ref{bkg_fit_results} with the fits shown in Fig.~\ref{bkg_spectra}. We found that the spectra between the three background regions were consistent with each other. Even though we included a 0.1~keV thermal plasma model to account for any LHB contribution, this component tended to 0 in each region. This is due to our restricted energy range of 0.4--7~keV, making our fits somewhat insensitive to this component. The defined background regions are ideal for some of our extended emission regions, but not all. In particular, Regions 4, 6, and 7 are not located adjacent to any (see Fig.~\ref{contour_plot}). This is because of our background selection criteria described above. To account for the background in these regions and given that the NE, NW, and SE backgrounds are consistent with each other, we generated a master background by fitting the spectra of all background regions simultaneously in XSPEC, the results of which are given in Table~\ref{bkg_fit_results} (Master) with the fits shown in Fig.~\ref{bkg_spectra}, bottom-right. These best fit parameters were used to fix the astrophysical background contribution in the extended Regions 4, 6, and 7.

\begin{table*}[!h]
\caption{Spectral fit results for thermal components background regions.}
\begin{center}
\label{bkg_fit_results}
\begin{tabular}{lllllll}
\hline
\hline
Region & $kT_{cool}$ & $K_{cool}$ & $kT_{hot}$ & $K_{hot}$ & $\chi^{2}_{\nu}$ (dof)\\
 & (keV)  & ($10^{-6}$~cm$^{-2}$~arcmin$^{-2}$) & (keV) & ($10^{-6}$~cm$^{-2}$~arcmin$^{-2}$) &  \\
\hline
NE  & 0.18 (0.17--0.19) & 2.25 (2.12--2.38)  & 0.77 (0.69--0.85) & 0.28 (0.21--0.35) & 1.14 (5994) \\ 
NW  & 0.18 (0.17--0.19) & 2.41 (2.18--2.62)  & 0.91 (0.84--1.03) & 0.38 (0.31--0.42) & 1.11 (5508) \\ 
SE  & 0.17 (0.15--0.20) & 4.02 (3.33--4.87)  & 0.70 (0.59--0.79) & 0.30 (0.20--0.34) & 1.12 (4790) \\ 
Master  & 0.18 (0.17--0.19) & 2.32 (2.24--2.40)  & 0.85 (0.81--0.90) & 0.37 (0.34--0.40) & 1.33 (20621) \\ 

\hline
\end{tabular}
\tablefoot{The numbers in parentheses are the 90\% confidence intervals. }
\end{center}
\end{table*}%

\subsubsection{Unresolved sources}\label{unresolved}

In addition to the background and extended emission contributions, the extended region spectra will be contaminated by emission from unresolved point sources. This will comprise a multitude of X-ray sources which are intrinsically weak such as CVs and ABs, or the faint end of the XRBs, supernova remnant (SNR), and supersoft source (SSS) populations. It is important to account for these components in the spectral analysis to isolate the truly diffuse emission contribution. In the following we describe how we account for these in the spectral fits.

\begin{figure*}[t]
\begin{center}
\resizebox{\hsize}{!}{\includegraphics[trim= 0.0cm 0.0cm 0.0cm 0.0cm, clip=true, angle=0]{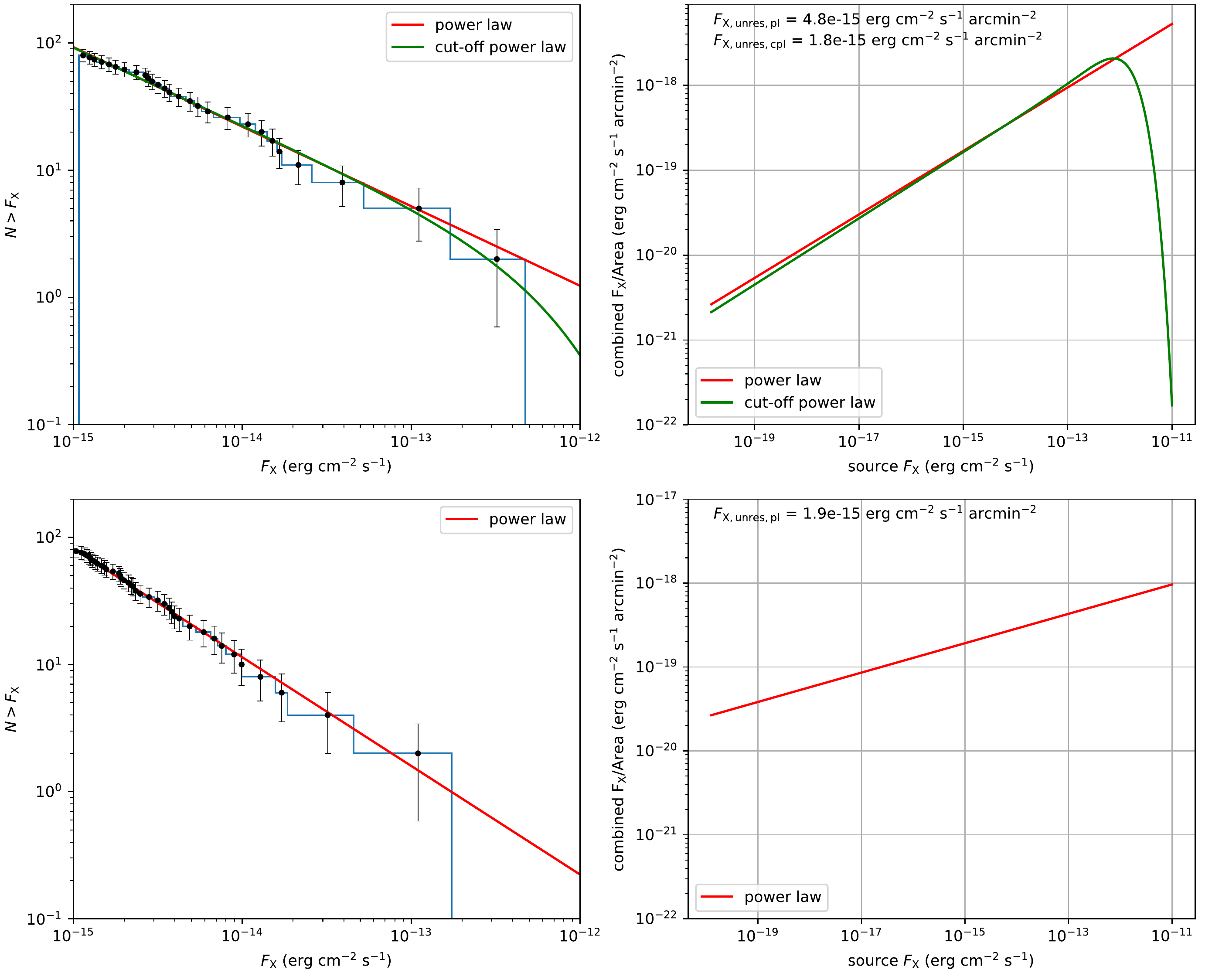}}
\caption{X-ray flux distributions of \gal\ sources detected in the southern (top row) and northern (bottom row) deep fields in the northern disk. The left panels show the X-ray flux functions, along with power law and cut-off power law fits. The right panels show the combined flux distribution expected from sources in flux bins, normalised for the deep field areas. The unresolved flux per unit area is determined by integrating over this distribution and subtracting the flux of detected sources. }
\label{unresolved_flux}
\end{center}
\end{figure*}

For the unresolved, though intrinsically bright X-ray sources, such as the XRBs, SNRs, and SSSs, we estimated the unresolved flux by constructing X-ray flux functions using the source catalogues of Paper~I for each of the deep fields in the northern disk (see Fig.~\ref{unresolved_flux}). Only sources with confirmed source types were included. To estimate the unresolved flux we first fitted the X-ray flux functions with either a power law or cut-off power law model (see Fig.~\ref{unresolved_flux}, left panel of both rows). The cut-off power law provides a slightly better fit to sources in the southern field of the northern disk, closer to the nuclear region of \gal. We therefore adopted this model to describe the northern sources, though the power law fit is included in Fig.~\ref{unresolved_flux} for completeness. For the northern field, the cut-off power law fit was indiscernible from the normal power law so we only consider the latter. Using these fits we then determined the combined flux per unit area from sources in each flux bin, normalised for the deep field area (Fig.~\ref{unresolved_flux}, right panels). We integrated over this distribution to determine the total flux per unit area expected from the entire source population and subtracted the resolved component to obtain the unresolved flux per unit area. We found the unresolved populations for the southern and northern deep fields to be $1.8\times10^{-15}$~erg~cm$^{-2}$~s$^{-1}$~arcmin$^{-2}$ and $1.9\times10^{-15}$~erg~cm$^{-2}$~s$^{-1}$~arcmin$^{-2}$, respectively. These values were then used to estimate the normalisations for the unresolved model component in spectral fits to the extended regions. However, we also require a spectral shape for this component.

\begin{figure}[t]
\begin{center}
\resizebox{\hsize}{!}{\includegraphics[trim= 0.0cm 0cm 0cm 0cm, clip=true, angle=0]{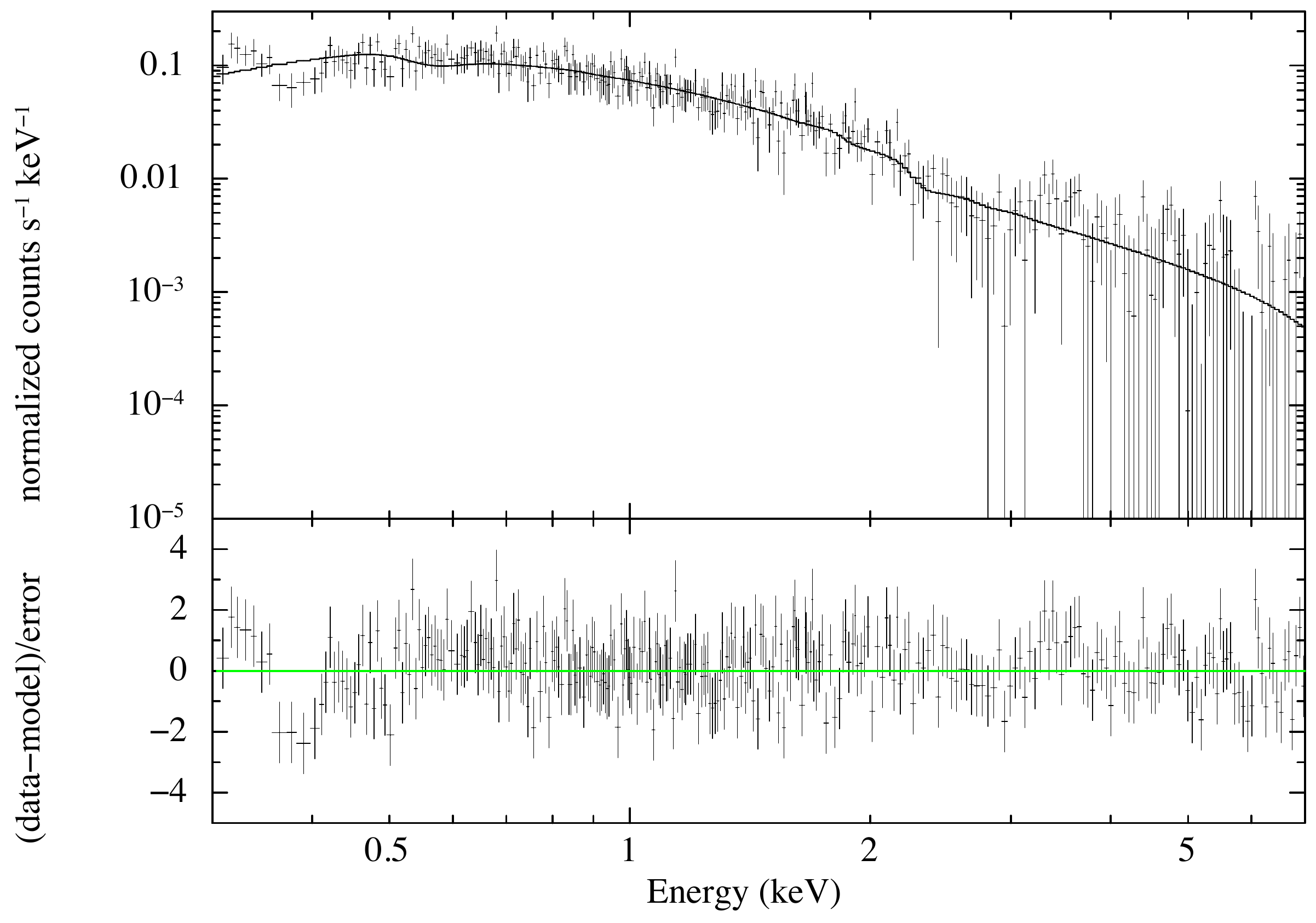}}
\caption{Combined faint source spectrum, fitted with an absorbed power law model with best fit $N_{H} = 0.96~(0.74-1.2)\times10^{21}$~cm$^{-2}$ and $\Gamma=2.37~(2.22-2.53)$ giving a reduced $\chi^{2}=1.07$ for 662 degrees of freedom}
\label{faint_source_spectrum}
\end{center}
\end{figure}

To estimate the spectral shape, we extracted spectra, backgrounds and response files from each of the sources used to construct the X-ray flux functions in Fig.~\ref{unresolved_flux} and combined them using the SAS task \texttt{epicspeccombine}. Source regions were the same as the source exclusion regions described in Sect.~\ref{x-ray_background} and background regions were defined using annuli with inner and outer radii of 1.1 and 2.0 times the source extraction radius, respectively. We omitted very bright sources to prevent them from dominating the combined spectrum, setting a source cutoff of 500 counts per source. The combined faint source spectrum is shown in Fig.~\ref{faint_source_spectrum}. We fitted the spectrum with several spectral models but found that a simple absorbed power law was sufficient with $N_{H} = 0.96~(0.74-1.2)\times10^{21}$~cm$^{-2}$ and $\Gamma=2.37~(2.22-2.53)$ giving a reduced $\chi^{2}=1.07$ for 662 degrees of freedom. Based on the best fit spectrum and expected contribution of the unresolved sources, we determined model normalisations for regions in the southern and northern fields of $K=5.86\times10^{-7}$~photons~keV$^{-1}$~cm$^{2}$~s$^{-1}$~arcmin$^{-2}$ and $K=5.55\times10^{-7}$~photons~keV$^{-1}$~cm$^{2}$~s$^{-1}$~arcmin$^{-2}$ at 1~keV, respectively. These spectral parameters were frozen in the fits to the extended region spectra.

We cannot apply the same method to determine the contribution from the intrinsically weak source population since none were detected in Paper~I. These sources are associated with the older stellar population in \gal. \citet{Revnivtsev2007, Revnivtsev2008} demonstrated that the X-ray luminosity per unit stellar mass due to weak X-ray sources (CVs and ABs) associated with the old stellar population in the solar vicinity \citep{Sazonov2006} is compatible with that observed in other galaxies. Therefore, we can determine the expected X-ray luminosity from the unresolved
CVs and ABs in the $>1$~Gyr population using the mass or, equivalently, luminosity of the
stellar population, which can be estimated using NIR data, specifically the K$_{\rm{s}}$-band. We determined the combined K$_{\rm{s}}$-band luminosity of our extended regions using the 2MASS Large
Galaxy Atlas \citep{Jarrett2003}\footnote{See \burl{http://irsa.ipac.caltech.edu/applications/2MASS/LGA/}}. We first identified and excised foreground stars from the images. We initially obtained estimates for each extended region. However, it became clear during the spectral analysis that this contribution was most significant in Regions 5, 6, and 7, but remained small when compared to both the background and diffuse components. Nevertheless, we included a model to account for unresovled CVs and ABs in our models for these three regions. We determined $K_{s} = $9.6, 7.1, and 4.7 for Regions 5, 6, and 7, respectively. Converting these into a luminosities yields $L_{K_{s}} = 1.7 \times 10^{7}$~L$_{\sun}$ for Region~5, $L_{K_{s}} = 1.9 \times 10^{8}$~L$_{\odot}$ for Region~6, and $L_{K_{s}} = 1.6 \times 10^{9}$~L$_{\odot}$ for Region~7. We then used the relation
$L_{X,0.5-2 \mathrm{keV}}/L_{K_{s}} = (5.9 \pm 2.5) \times 10^{27}$~erg~s$^{-1}$~L$^{-1}_{K_{s} ,L_{\sun}}$ \citep{Revnivtsev2008} and adjusted for the X-ray energy range using their emission model, a \texttt{mekal} model with $kT = 0.5$~keV and solar abundance, and a power law with
$\Gamma = 1.9$, with the ratio of component luminosities equal to 2.03, to determine an expected contribution to the X-ray emission of $L_{X, 0.3-10 \mathrm{keV}} = (1.1\pm0.4)\times10^{35}$~erg~s$^{-1}$ for Region 5, $L_{X, 0.3-10 \mathrm{keV}} = (1.1\pm0.5)\times10^{36}$~erg~s$^{-1}$ for Region 6, and $L_{X, 0.3-10 \mathrm{keV}} = (9.8\pm0.4)\times10^{36}$~erg~s$^{-1}$ for Region 7. We then converted these values to flux and normalised for area. The resulting values were used with the \texttt{mekal+powerlaw} model of \citep{Revnivtsev2008} to determine the model normalisations $K_{\rm{mekal}}=7.61\times10^{-10}$~cm$^{-5}$~arcmin$^{-2}$, $K_{\rm{mekal}}=3.40\times10^{-8}$~cm$^{-5}$~arcmin$^{-2}$, and $K_{\rm{mekal}}=9.16\times10^{-8}$~cm$^{-5}$~arcmin$^{-2}$ for Regions 5, 6, and 7, respectively. The power law normalisations were tied to the \texttt{mekal} normalisation so that their luminosity ratios were 2.03, as per \citet{Revnivtsev2008}.

Overall, the unresolved point sources were found to contribute significantly to the observed flux from the extended emission regions. From the spectral fits described in Section~\ref{spec_analysis}, we estimated that point sources account for $\sim7$~\% of the flux from Region~1, $\sim$14~\% from Region~2, $\sim$34~\% from Region~3, $\sim$35~\% from Region~4, $\sim$42~\% from Region~5, $\sim$59~\% from Region~6, and $\sim$71~\% from Region~7. The intrinsically weak sources contributed $<1$~\%, $\sim2$~\%, and $\sim7$~\% of the flux to Regions 5, 6, and 7, respectively. 

\subsubsection{\gal\ emission}
We expect any emission from the hot ISM plasma in \gal\ to be dominated by thermal processes. Therefore, we included thermal plasma models in the spectral fits absorbed by foreground Galactic and \gal\ material on top of the background components discussed in the preceding subsections. To fit the emission we proceeded through the following steps: we first tried a single-temperature thermal plasma in collisional ionisation equilibrium (CIE), namely the \apec\ model in XSPEC; if a good fit to the data was found we tried a non-equilibrium ionisation (NEI) model \citep[\nei\ in XSPEC,][]{Borkowski2001} to provide constraints on the ionisation timescale ($\tau$)\footnote{Plasmas with $\tau\gtrsim10^{12}$~s~cm$^{-3}$ indicate CIE conditions} of the plasma; if the \apec\ fit was bad, we tried the \nei\ model to test for a non-equilibrium plasma and two temperature models (combinations of \apec\ and \nei) in case the emission comprised two distinct thermal components, though the latter was never required. The extended emission being studied is very faint in nature and, even with the deep observations of the northern disk, the spectral statistics did not allow for any more complex models with variable abundance.

\subsubsection{Neighbouring region backgrounds}
It is clear from Fig.~\ref{contour_plot} (right) that there is some overlap of the larger extended emission regions with the more localised emission association with the young stellar populations, particularly along the dust ring. Namely, the extended emission in Region~5 extends across and likely contributes to emission in Regions~2 and 3. The extended emission in Region~7 is similarly affected by Region~4. To account for this we included a spectral component for Region 5 in fits to Regions~2 and 3, and from Region~7 in fits to Region~4. The contributions of these components were fixed to the best fits in Regions~5 and 7 and normalised per unit area.

\section{Ancillary data}
To compare the morphology and properties of the hot ISM plasma in \gal\ to the stellar and colder ISM components of the galaxy, we made use of multi-wavelength archival data presented in the literature, which we describe below.

\subsection{UV}
The {\it Galaxy Evolution Explorer} (\galex) satellite has surveyed \gal, obtaining images in the near-UV (NUV) and far-UV (FUV), the analysis of which is presented in \citet{Thilker2005}. The angular resolution  of \galex\ allows for the separation of individual young stellar clusters hosting massive stars from each other, their environments, and the diffuse UV emission in the galaxy. Therefore, these observations provide a tracer of the populations of massive stars driving the dynamics of the ISM. We obtained the \galex\ exposures from the Mikulski Archive for Space Telescopes\footnote{\burl{http://galex.stsci.edu/GR6/}}. We mosaicked the NUV and FUV fields using the Montage 
tool\footnote{\burl{http://montage.ipac.caltech.edu/}}.

\subsection{Optical}
\gal\ was observed as part of the Local Group Galaxy Survey (LGGS), performed using the {\it Kitt Peak National Observatory} 4~m telescope \citep{Massey2006}. The study covered 2.2 deg$^{2}$ along the major axis of M31 and includes all massive star forming regions. Broadband images were obtained in $U$, $B$. $V$, $R$, and $I$, as well as \ha, \sii, and \oiii\ narrow band, emission line images. The warm ISM ($10^{4}~K$) of \gal\ is revealed in \ha. Therefore, we downloaded both the bias-corrected and flat-fielded \ha\ and $R$-band images from the LGGS archive\footnote{\burl{www.lowell.edu/users/massey/lgsurvey.html}} in order to produce continuum subtracted \ha\ images to reveal the full extent of the faint diffuse warm ISM of \gal. The continuum subtraction was performed on the individual fields of the LGGS survey (10 in total) before constructing a mosaic of these fields. The $R$-band image was scaled to the \ha\ image in each field by first identifying stellar sources in each using the \texttt{DAOStarFinder} task \citep[based on \texttt{DAOFind},][]{Stetson1987} contained in the Astropy \citep{Astropy2013} affiliated package Photutils \citep{photutils2016}, determining the scaling relation based on their fluxes, then scaling and subtracting the continuum image from the emission line image. The resulting continuum subtracted fields were mosaicked using the Montage tool. 

\subsection{\hi}
Neutral H in the \gal\ ISM can be traced using the 21~cm emission line. Therefore, we make use of the high-resolution \hi\ cube presented in \citet{Braun2009}, obtained using the {\it Westerbork Synthesis Radio Telescope} and the {\it Green Bank Telescope}. These data reveal the atomic gas component in \gal\ with a $\sim15\arcsec$ spatial and 2.3~km~s$^{-1}$ velocity resolution in a $7\times3.5$~deg$^2$ region.

\subsubsection{Infrared and sub-mm}
The cold ISM of \gal\ can be revealed by infrared (IR) emission. We used data from the {\it Spitzer Space Telescope} \citep{Werner2004} and the {\it Herschel Space Observatory} \citep{Pilbratt2010}. The 24~$\mu$m band of the Multiband Imaging Photometer \citep[MIPS,][]{Rieke2004} provides us with a picture of the stochastically and thermally heated dust in the ISM, and is presented in \citet{Gordon2006}. We obtained these data from the IPAC Infrared Science Archive\footnote{\burl{http://irsa.ipac.caltech.edu/data/SPITZER/}}. Colder dust ($\sim$10~K) requires observations longward of 200~$\mu$m, in the wavelength range of the \herschel\ Spectral and Photometric Imaging Receiver \citep[SPIRE,][]{Griffin2010}. Observations of \gal\ with \herschel\ SPIRE are reported as part of the \herschel\ Exploitation of Local Galaxy Andromeda (HELGA) survey by \citet{Fritz2012}. For our study, we make use of the SPIRE 250~$\mu$m data to trace the cold dust morphology of \gal. 

\subsubsection{Panchromatic Hubble Andromeda Treasury}
The Panchromatic Hubble Andromeda Treasury \citep[PHAT;][]{Dalcanton2012} is a 
{\em Hubble} Space Telescope (HST) survey which observed one-third of the star-forming disk of \gal\ in six filters, from the UV to the NIR. The survey provided luminosities and colours for over 100 million individual stars. We made use of the stellar photometry collected in this survey \citep[F475W and F814W filters,][]{2018ApJS..239...13W} for the study of the stellar populations in our regions of interest.

\section{Results}
\label{results}

\begin{table*}
\caption{Spectral fit results to the extended emission regions in the northern disk of \gal.}
\begin{center}
\label{spectral_fit_results}
\begin{tabular}{llllllll}
\hline
\hline
Model & $N_{\rm{H}}$ & $kT_{1}$ & $\tau_{1}$ & $EM^{\rm{a}}_{1}$  & $\chi^{2}_{\nu}$ (dof) & log $F_{\rm{X}}^{(b)}$ & log $L_{\rm{X}}^{(c)}$  \\

 & ($10^{22}$~cm$^{-2}$) & (keV) & ($10^{11}$~s~cm$^{-3}$) & ($10^{60}$~cm$^{-3}$) & & (erg~cm$^{-2}$~s$^{-1}$) & (erg~s$^{-1}$)  \\
\hline
\multicolumn{8}{c}{\textbf{Region 1}} \\
\texttt{apec} & 0.62 (0.39--0.76) & 0.20 (0.18--0.24) & --  & 6.29 (2.69--14.08) &  0.99 (4329) & -13.37 & 37.85 \\   
\texttt{nei} & 0.69 (0.66--0.89) & 0.19 (0.18--0.46) & 30.09 ($>4.06$)  & 8.15 (4.24--19.23)  &  0.99 (4328) & -13.38 & 37.85 \\ 
\multicolumn{8}{c}{\textbf{Region 2}} \\
\texttt{apec} & 0.07 (0.02--0.10) & 0.60 (0.56--0.65) & --  & 0.14 (0.11--0.15)  &  0.99 (4994) & -13.60 & 36.44 \\ 
\multicolumn{8}{c}{\textbf{Region 3}} \\
\texttt{apec} & 0.08 (0.05--0.22) & 0.33 (0.24--0.67) & --  & 0.04 (0.02--0.12)  &  1.06 (1099) & -14.39 & 35.76 \\ 
\multicolumn{8}{c}{\textbf{Region 4}} \\
\texttt{apec} & ($<0.05$) & 0.20 (0.18--0.21) & --  & 0.36 (0.32--0.49) &  0.98 (3086) & -13.55 & 36.60 \\ 
\multicolumn{8}{c}{\textbf{Region 5}} \\
\texttt{apec} & 0.18 ($<0.34$) & 0.10 (0.07--0.15) & --  & 10.36 (3.01--52.78)  &  1.11 (6147) & -13.53 & 37.41 \\ 
\multicolumn{8}{c}{\textbf{Region 6}} \\
\texttt{apec} & $<0.39$ & 0.19 (0.18--0.22) & --  & 1.43 (1.03--6.41) &  1.01 (4486) & -13.66 & 36.50 \\ 
\texttt{nei} & 0.11 ($<0.40$) & 0.18 (0.17--0.21) & $>9.27$  & 2.74 (0.90--12.61)  &  1.01 (4485) & -13.64 & 36.78 \\ 
\multicolumn{8}{c}{\textbf{Region 7}} \\
\texttt{apec} & 0.20 (0.07--0.34) & 0.94 (0.86--0.99) & --  & 0.33 (0.24--0.47) &  1.13 (5204) & -13.30 & 36.80 \\
\hline

\end{tabular}
\tablefoot{The numbers in parentheses are the 90\% confidence intervals.
\tablefoottext{a}{Emission measure, calculated from the model normalisation ($K$) as $EM = (K~4\pi d^{2})/10^{-14}$ where $d$ is the distance to \gal\ in cm.}
\tablefoottext{b}{0.3-8~keV absorbed X-ray flux.}
\tablefoottext{c}{0.3-8~keV unabsorbed X-ray luminosity.}
}
\end{center}
\end{table*}%

\subsection{Spectral analysis}
\label{spec_analysis}
The spectral analyses of each individual extended emission regions are presented in the following sub-sections. For clarity, we only show spectral plots of the EPIC spectrum with the most counts for each, along with additive components of interest (Figs.~\ref{spectral_fits1_rev} and \ref{spectral_fits2_rev}). Plots showing the simultaneous fits to each observational and FWC spectrum, along with all additive components, are given in Appendix~B for completeness (Figs.~\ref{spectral_fits1} and \ref{spectral_fits2}).

\subsubsection{Region 1}
\label{reg1_fits}
\begin{figure*}
\begin{center}
\resizebox{6.3in}{!}{\includegraphics[trim= 0cm 0cm 0cm 0cm, clip=true, angle=0]{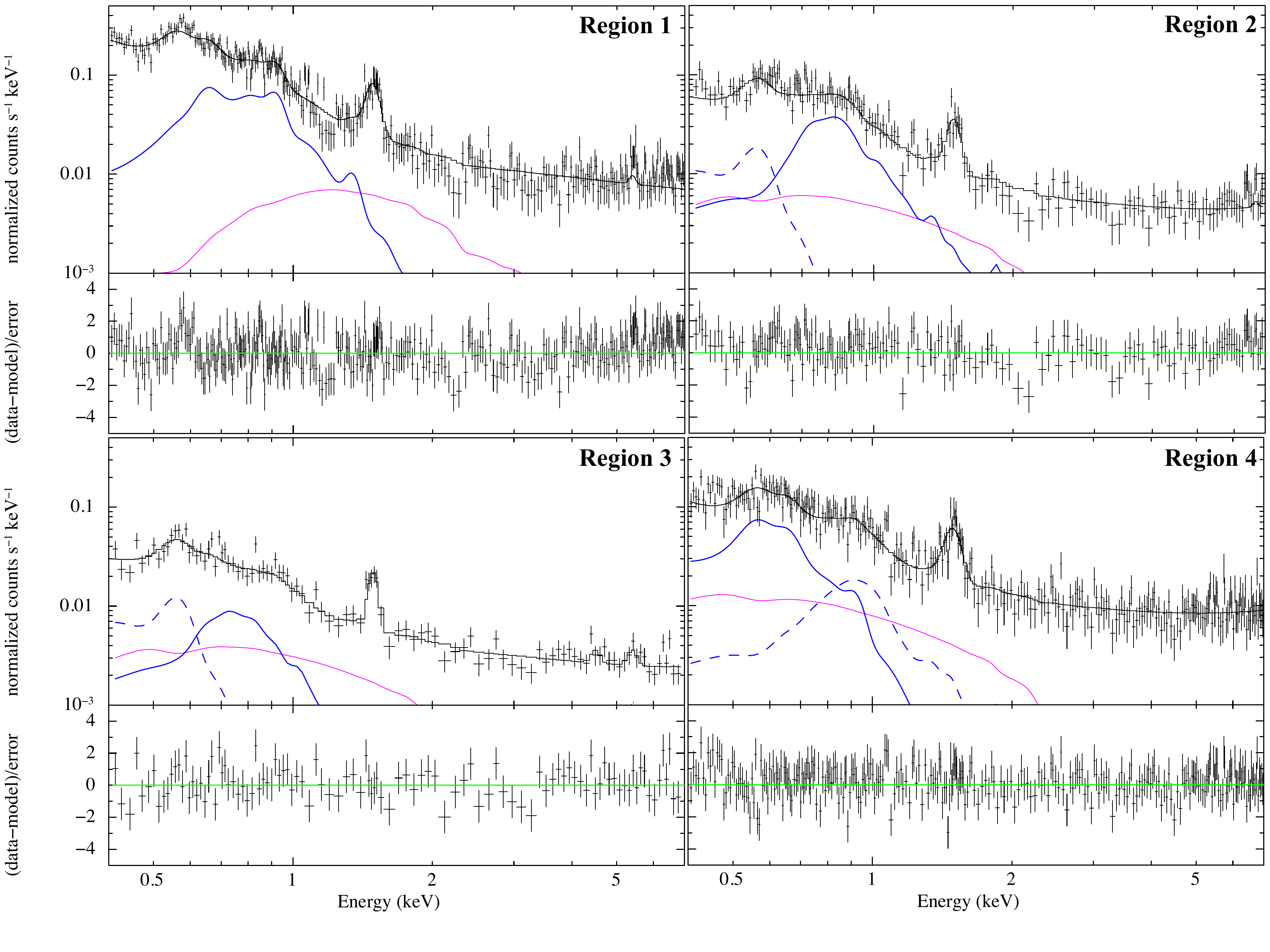}}
\caption{\xmm\ spectral fits of our extended emission Regions~1 through 4. Each panel shows the EPIC spectrum with the most counts along with additive components of interest for clarity. The black data points and associated solid lines are the observational spectra and best fit model, respectively. The magenta lines show the unresolved SNR, XRB, and SSS component (see Sect.~\ref{x-ray_background} for descriptions), blue solid lines represent the \gal\ ISM component with the dashed blue lines indicating the contamination of smaller regions by larger extended emission regions, that is Region~5 components in Regions~2 and 3, and Region~7 component in Region~4 (see Fig.~\ref{multiwave_images}). Best fit parameters are given in Table~\ref{spectral_fit_results}. More detailed plots are presented in Fig.~\ref{spectral_fits1}.}
\label{spectral_fits1_rev}
\end{center}
\end{figure*}

We found that the emission over the background components in Region 1 could be fitted with a single \texttt{apec} model with $kT=0.20~(0.18-0.24)$~keV with \nh~$=0.62~(0.39-0.76)\times10^{22}$~cm$^{-2}$ giving a reduced $\chi^{2}$ ($\chi^{2}_{\nu}) = 0.99$. We also tried the \texttt{nei} model which provided a similarly good fit and found good agreement with the \apec\ model, with the additional free parameter $\tau>4.06\times10^{11}$~s~cm$^{-3}$ suggesting a plasma close to or in CIE. The fit results are given in Table~\ref{spectral_fit_results}, with the spectra and \nei\ fits shown in Fig.~\ref{spectral_fits1_rev}, top left.

\subsubsection{Region 2}
\label{reg2_fits}
We fitted the extended emission with a single \texttt{apec} model and found $kT=0.60~(0.56-0.65)$ keV with \nh$=0.07~(0.02-0.10)\times10^{22}$~cm$^{-2}$ giving a $\chi^{2}_{\nu} = 0.99$ (see Table~\ref{spectral_fit_results}). We also tried the \texttt{nei} model, however the $\tau$ parameter was not well constrained rendering the results difficult to interpret. The spectra and \apec\ fits are shown in Fig.~\ref{spectral_fits1_rev}, top right.

\subsubsection{Region 3}
\label{reg3_fits}
Region~3 was well fitted by a single \texttt{apec} model with $kT=0.33~(0.24-0.67)$ keV with \nh$=0.08~(0.05-0.22)\times10^{22}$~cm$^{-2}$ giving a $\chi^{2}_{\nu} = 1.06$ (see Table~\ref{spectral_fit_results}). As with Region 2, we also tried the \texttt{nei} model but the $\tau$ parameter was not well constrained so the model does not provide additional insight into the plasma conditions. The spectra and \apec\ fits are shown in Fig.~\ref{spectral_fits1_rev}, bottom left.

\subsubsection{Region 4}
\label{reg4_fits}
We fitted the spectra of Region~4 with a single \apec\ model which provided a good fit to the data ($\chi^{2}_{\nu} = 0.98$) with a best-fit plasma temperature of $0.20~(0.18-0.21$)~keV and \nh$<0.04\times10^{22}$~cm$^{-2}$ (see Table~\ref{spectral_fit_results}). No further constraints on the plasma conditions were provided using the \nei\ model. The spectra and \apec\ fits are shown in Fig.~\ref{spectral_fits1_rev}, bottom right.

\begin{figure*}
\begin{center}
\resizebox{7in}{!}{\includegraphics[trim= 0cm 0cm 0cm 0cm, clip=true, angle=0]{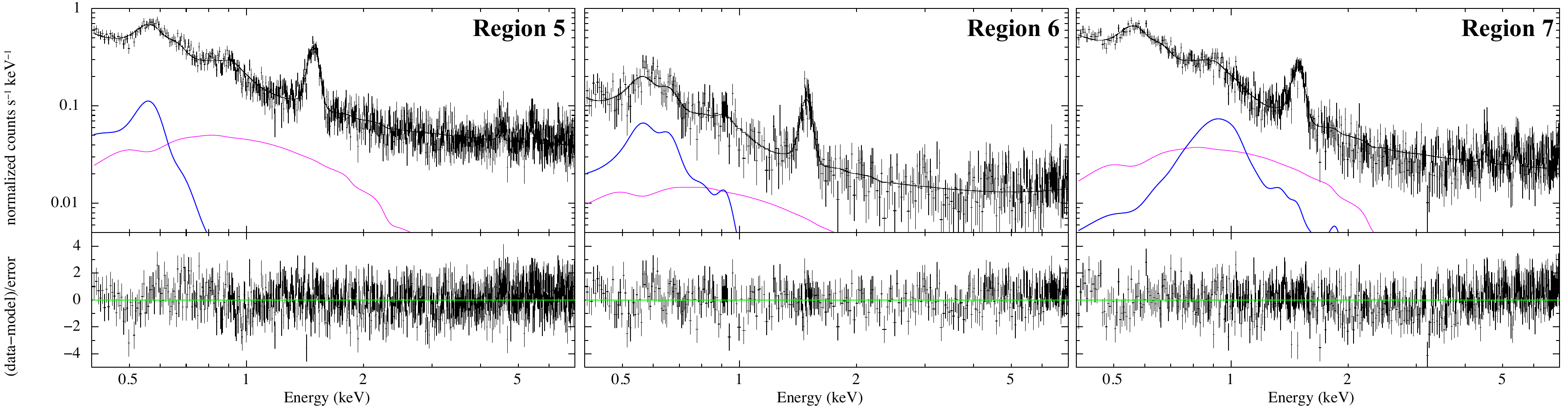}}
\caption{Same as Fig.~\ref{spectral_fits1_rev} for Regions 5, 6, and 7, arranged from left to right. Best fit parameters are given in Table~\ref{spectral_fit_results}. More detailed plots are presented in Fig.~\ref{spectral_fits2}.}
\label{spectral_fits2_rev}
\end{center}
\end{figure*}

\subsubsection{Region 5}
The extended emission in Region~5 was well characterised by a single \apec\ model ($\chi^{2}_{\nu} = 1.11$) with a best-fit plasma temperature of 0.10~($0.07-0.15$)~keV and \nh$=0.18~(<0.34)\times10^{22}$~cm$^{-2}$ (see Table~\ref{spectral_fit_results}). The faintness of the extended emission in Region~5 over the background meant the \nei\ model did not provide further information on the plasma conditions. The spectra and \apec\ fits are shown in Fig.~\ref{spectral_fits2_rev}, left.

\subsubsection{Region 6}
A single \apec\ model provided a good fit to the emission in Region 6 with $\chi^{2}_{\nu} = 1.01$ for $kT=0.19~(0.18-0.22)$~keV and \nh$<0.39\times10^{22}$~cm$^{-2}$ (see Table~\ref{spectral_fit_results}). We also applied the \nei\ model which provided a constraint on the ionisation parameter with $\tau>9.27\times10^{11}$~s~cm$^{-3}$ suggesting a plasma in collisional ionisation equilibrium. The spectra and \nei\ fits are shown in Fig.~\ref{spectral_fits2_rev}, middle.

\subsubsection{Region 7}
Again, the extended emission in Region~7 was well accounted for by a single \apec\ model ($\chi^{2}_{\nu} = 1.13$) with $kT=0.94~(0.86-1.99)$~keV and \nh$=0.20~(0.07-0.34)\times10^{22}$~cm$^{-2}$ (see Table~\ref{spectral_fit_results}). We also applied the \nei\ model, though the fits did not yield any further information. The spectra and \apec\ fits are shown in Fig.~\ref{spectral_fits2_rev}, right.

\section{Discussion}
\label{discussion}
\subsection{Multi-wavelength view}
\label{mwm}
\begin{figure*}
\begin{center}
\resizebox{\hsize}{!}{\includegraphics[trim= 0.0cm 0cm 0cm 0cm, clip=true, angle=0]{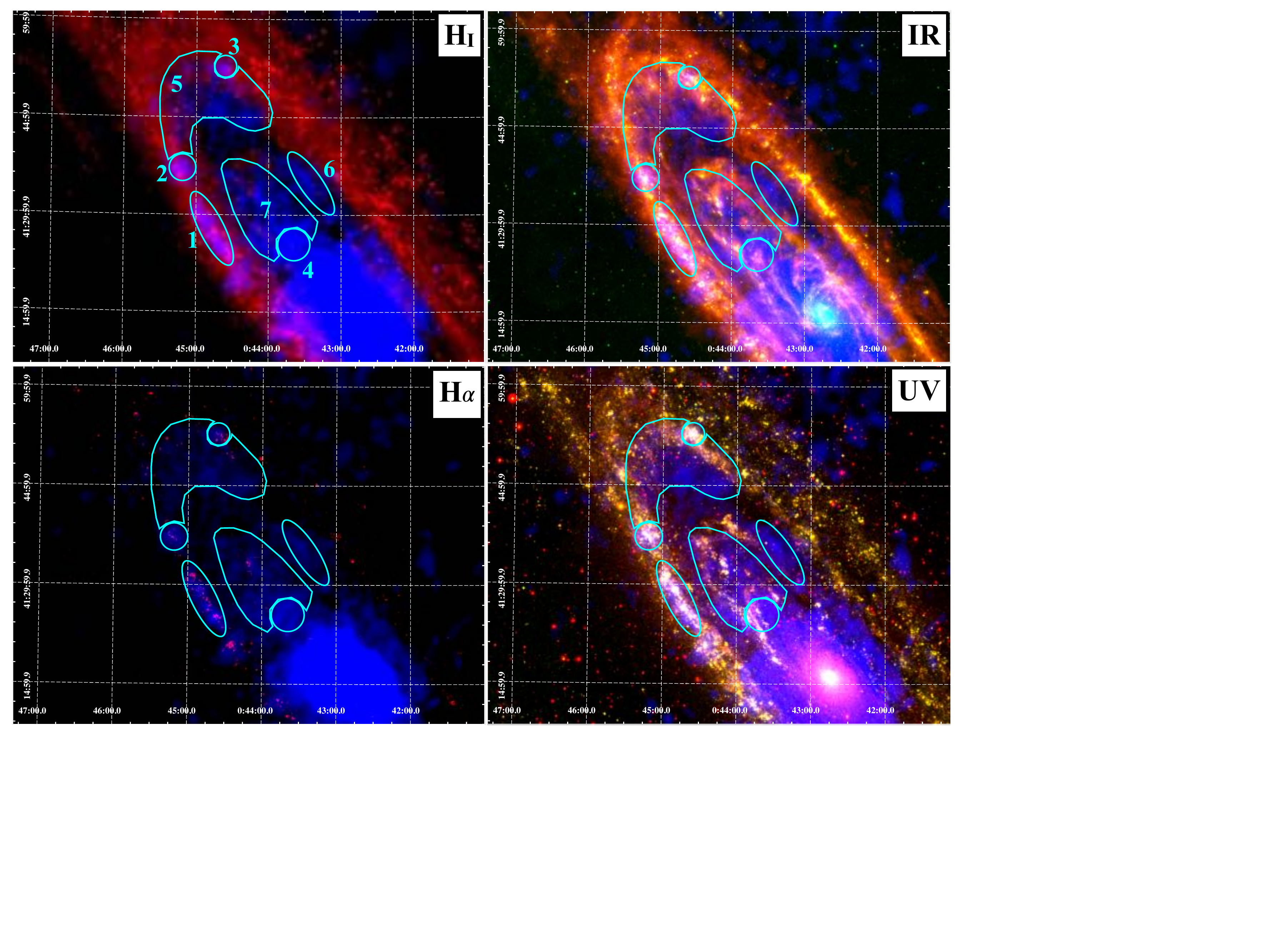}}
\caption{Multi-wavelength comparison of hot ISM plasma in the northern disk of \gal\ with high mass stellar and cooler ISM components. In each case, the 0.3-0.7~keV X-ray image is given in blue and excised point sources have been filled. Top-left shows the comparison with \hi\ (red). Top-right shows the comparison with the dust content, revealed in 250~$\mu$m (red) by \herschel\ SPIRE and 24~$\mu$m (green) by \spitzer\ MIPS. Bottom-left shows \ha\ (red) from the LGGS, which traces the warm ISM, and bottom-right shows the location of the massive stellar population, revealed in the NUV (red) and FUV (green) by \galex. In all panels, the extended X-ray emission regions from our spectral analysis are shown to guide the eye, with labels given in top-right only.}
\label{multiwave_images}
\end{center}
\end{figure*}

\begin{figure}
\begin{center}
\resizebox{\hsize}{!}{\includegraphics[trim= 0.0cm 0cm 0cm 0cm, clip=true, angle=0]{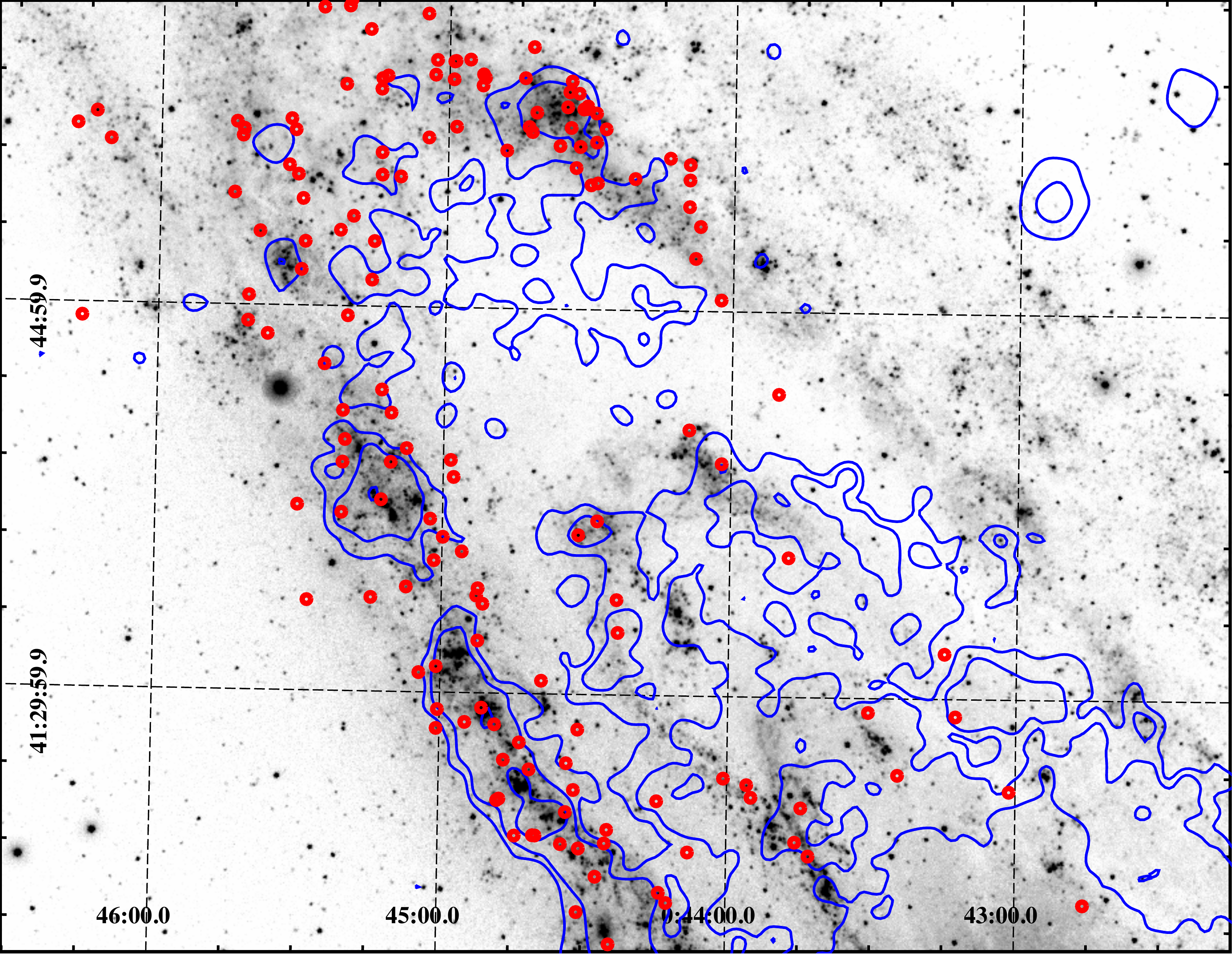}}
\caption{NUV image from Fig.~\ref{multiwave_images} bottom-right, with stellar clusters from \citet{Johnson2016} shown by the red dots. The clusters have mass greater than $10^{3}$~M$_{\sun}$ and ages less than 100~Myr. The 0.4--1.25~keV X-ray contours from Fig.~\ref{contour_plot} are also shown in blue to highlight the correlation between the stellar clusters and X-ray emission. No clusters are plotted in the 10~kpc ring north of the core as this region was not studied in \citet{Johnson2016}.}
\label{phat_clusters}
\end{center}
\end{figure}

The X-ray morphology is compared to \hi, IR, \ha, and UV views of the northern disk in Fig.~\ref{multiwave_images}. The multi-wavelength view of the northern disk is dominated by the 10~kpc star-forming ring \citep{Arp1964,Habing1984}. The prevalence of massive stars in the 10~kpc ring is clear from the \ha\ and FUV images (Fig.~\ref{multiwave_images}, bottom-left, bottom-right, and Fig.~\ref{phat_clusters}). The inner dust ring at $\sim5$~kpc also hosts massive stellar clusters (MSCs) with significant UV flux there (Fig.~\ref{multiwave_images}, top-right, bottom-right). 

The stellar cluster population of the northern disk of \gal\ has been catalogued using the PHAT data \citep{Johnson2012,Johnson2015}. Ages and masses for many of these have been determined by \citet{Johnson2016}. We used this catalogue to identify clusters with masses greater than $10^{3}$~M$_{\sun}$ with ages less than 100~Myr and plotted these over the NUV image, which is shown in Fig.~\ref{phat_clusters}. However, this study did not include the 10~kpc ring to the north and northwest of the \gal\ core so clusters in those regions are absent from Fig.~\ref{phat_clusters}. Nevertheless, there is an obvious correlation between the stellar cluster distribution and the extended X-ray emission, in particular along the 10~kpc ring to the northeast and east of the core, clear evidence of the link between the massive stellar population and the extended X-ray emission. 

We found that the brightest extended X-ray emission in the northern disk follows closely the brightest regions in \ha\ and UV. Therefore, we conclude that the populations of massive stars in these regions play and important role in the production of the X-ray emission. These regions are discussed in more detail in Sect.~\ref{msfrs}. Fainter extended X-ray emission was detected in the low density regions between the 10~kpc and 5~kpc rings and between the 5~kpc ring and the core. The emission appears anti-correlated with the dust structures (Fig.~\ref{multiwave_images}, top-right) and does not appear to be associated with MSCs with intense UV flux as for Regions~1--4. These regions are discussed in detail in \ref{large_scale}.

\subsection{Emission from MSCs}
\label{msfrs}
\subsubsection{Region 1}
\begin{sidewaysfigure*}[!h]
\begin{center}
\resizebox{\hsize}{!}{\includegraphics[trim= 0cm 0cm 0cm 0cm, clip=true, angle=0]{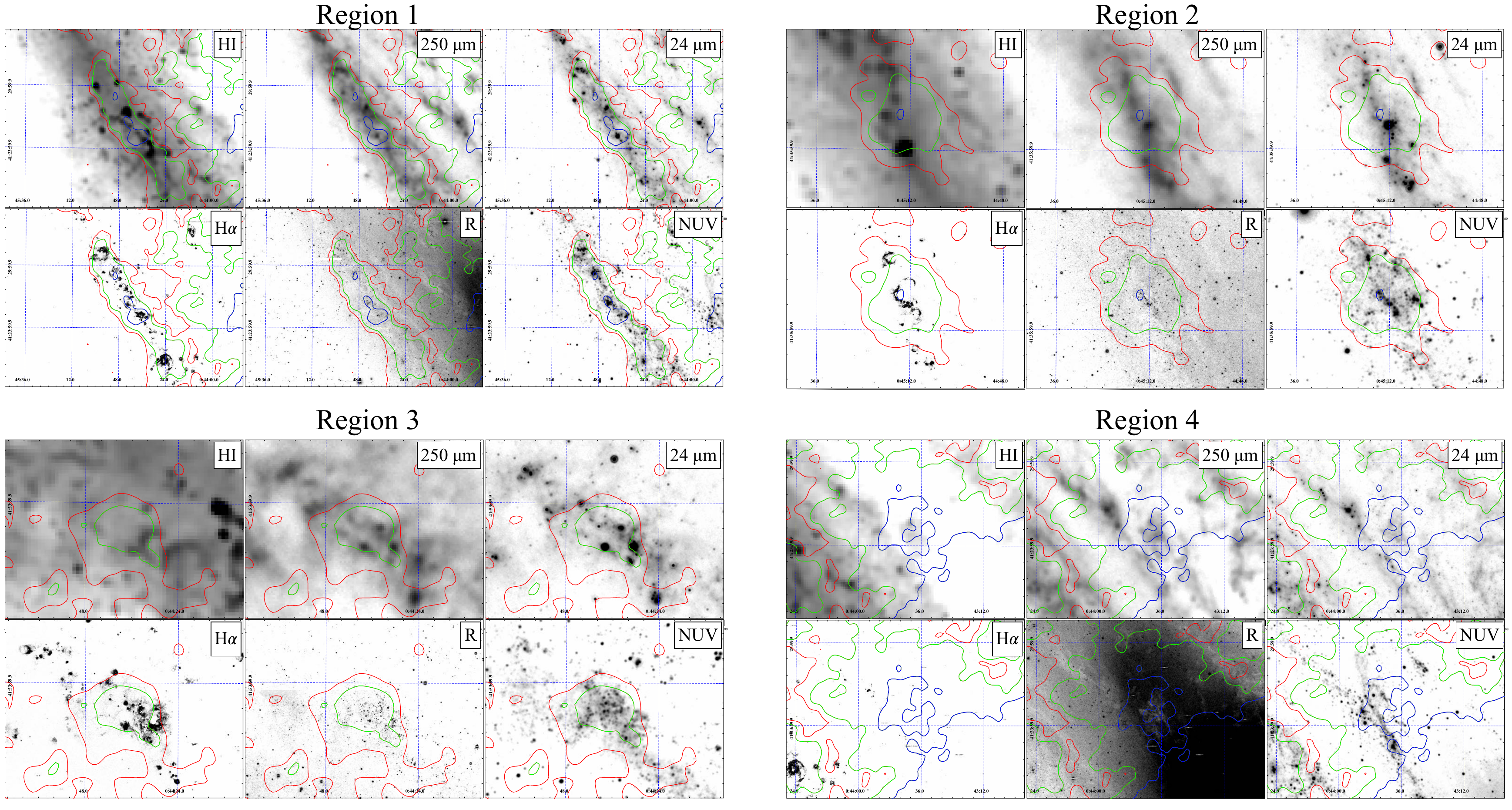}}
\caption{Close-up, multi-wavelength views of extended emission regions associated with populations of massive stars with the wavebands indicated in the panes. The red, green, and blue contours correspond to $5\sigma$, $10\sigma$, and $20\sigma$ above the average background level determined from regions exterior to the northern disk. The individual regions are indicated on the plot}
\label{multi_mw}
\end{center}
\end{sidewaysfigure*}

Region~1 spans what is probably the richest in massive stars region in the 10~kpc ring. Because of its extended nature along the ring itself, the extended X-ray emission likely results from several MSCs. This is evident from the multi-wavelength comparison shown in Fig.~\ref{multi_mw}, top-left, with the extended X-ray emission clearly corresponding to enhancements in the \ha\ and NUV images. Knots of colder material, shown in the \hi, 250~$\mu$m, and 24~$\mu$m images, appear distributed along the extent of Region~1, and could be sites of active star formation.

\subsubsection{Region 2}
The multi-wavelength view of Region~2 (Fig.~\ref{multi_mw}, top-right) shows that X-ray emission likely results from several MSCs. \citet{Lewis2015} have shown that this region exhibited a high star formation rate in the last 25~Myr. The brightest region of the X-ray emission is centred on a large \ha\ shell measuring $\sim300$~pc in diameter. The shell is bright on the eastern side though it appears broken in the west, possibly due to blowouts. Both the R-band and NUV images indicate an enhanced stellar density along the eastern edge of the shell, where the IR and \hi\ emission is highest. The density gradient in these regions could explain the observed dichotomous morphology of the \ha\ shell. This multi-wavelength picture suggests that the X-ray emission in Region~2 could originate in a large superbubble. The size of the \ha\ shell is rather large for a superbubble, which typically have diameters of $\sim10^{2}$~pc around young MSCs \citep[e.g.][]{Sasaki2011,Kavanagh2012,DeHorta2014}. It is possible that the \ha\ structure in Region~2 has been driven by more than one generation of MSCs, and is transitioning from a superbubble to a supergiant shell.

\subsubsection{Region 3}
Region~3 was the brightest region of X-ray emission along the northern-most arc of the 10~kpc ring.  The multi-wavelength view of Region~3 (Fig.~\ref{multi_mw}, bottom-left) shows significant \ha\ emission towards the western side. The R-band and NUV images show MSCs are present, mostly concentrated in the centre of the region, with some extension to the southwest into the \ha\ structures. This region was found to have a high star formation rate over the last 50~Myr \citep{Lewis2015}. A slight east-west density gradient is also seen in the \hi\ and IR images, which may explain the \ha\ morphology.

\subsubsection{Region 4}
The multi-wavelength view of Region~4 (see Fig.~\ref{multi_mw}, bottom-right) shows no \ha\ structure associated with the X-ray enhancement, though the NUV image suggests that MSCs are present. Interestingly, the R-band image shows a slight deficit of stars compared to the surrounding regions. A density enhancement in \hi, 250~$\mu$m, and 24~$\mu$m appears to be located in the foreground, absorbing the X-ray emission on the western side. This density enhancement may also explain the apparent deficit in stellar density seen in the R-band as resulting from foreground absorption.

\subsubsection{Mechanical input from MSCs}
\label{stellar_input}
As discussed earlier in this paper, the X-ray emission in Regions~1--4 most likely results from the mechanical input by massive stars and SNe in the MSCs of these regions. In this sub-section, we estimate the power injected into each region by the stellar populations using the PHAT photometric data with stellar evolution models and a population synthesis code. The results for each region are discussed below.

\begin{figure*}[!h]
\begin{center}
\resizebox{6.8in}{!}{\includegraphics[trim= 0cm 0cm 0cm 0cm, clip=true, angle=0]{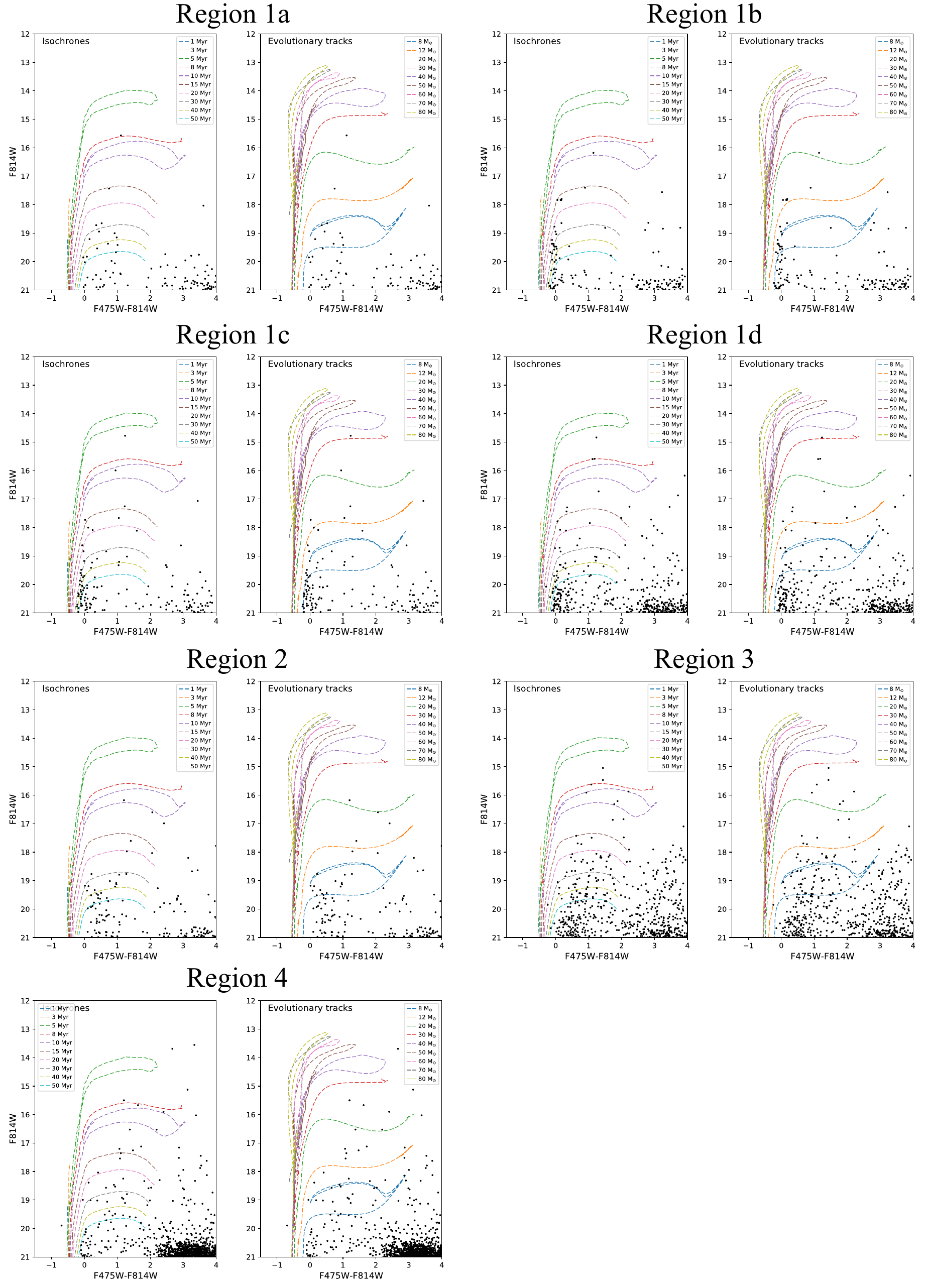}}
\caption{Colour-magnitude diagrams for the extended emission regions associated with MSCs. The data points are from PHAT photometry. Isochrones and evolutionary tracks derived from the PARSEC isochrone database are shown on the left and right, respectively.}
\label{all_cmds}
\end{center}
\end{figure*}

We constructed colour-magnitude diagrams for stars located within the regions using the F475W and F814W filters. To determine the properties of the stellar population we obtained stellar evolutionary tracks in the corresponding filters from the CMD~3.0 tool\footnote{See \burl{http://stev.oapd.inaf.it/cmd}}, based on the PARSEC isochrones \citep{Bressan2012,Chen2014,Chen2015,Tang2014}. We selected the Chabrier lognormal IMF \citep{Chabrier2001} for M$<$~1M$_{\sun}$ and the Salpeter IMF \citep{Salpeter1955} for M$>$1~M$_{\sun}$, and assumed Galactic abundance. We extracted isochrones of various ages and evolutionary tracks at various masses and overplotted these on the PHAT photometry for the stellar populations in each region, shown in Fig.~\ref{all_cmds}.

\begin{figure*}[!h]
\begin{center}
\resizebox{\hsize}{!}{\includegraphics[trim= 0cm 0cm 0cm 0cm, clip=true, angle=0]{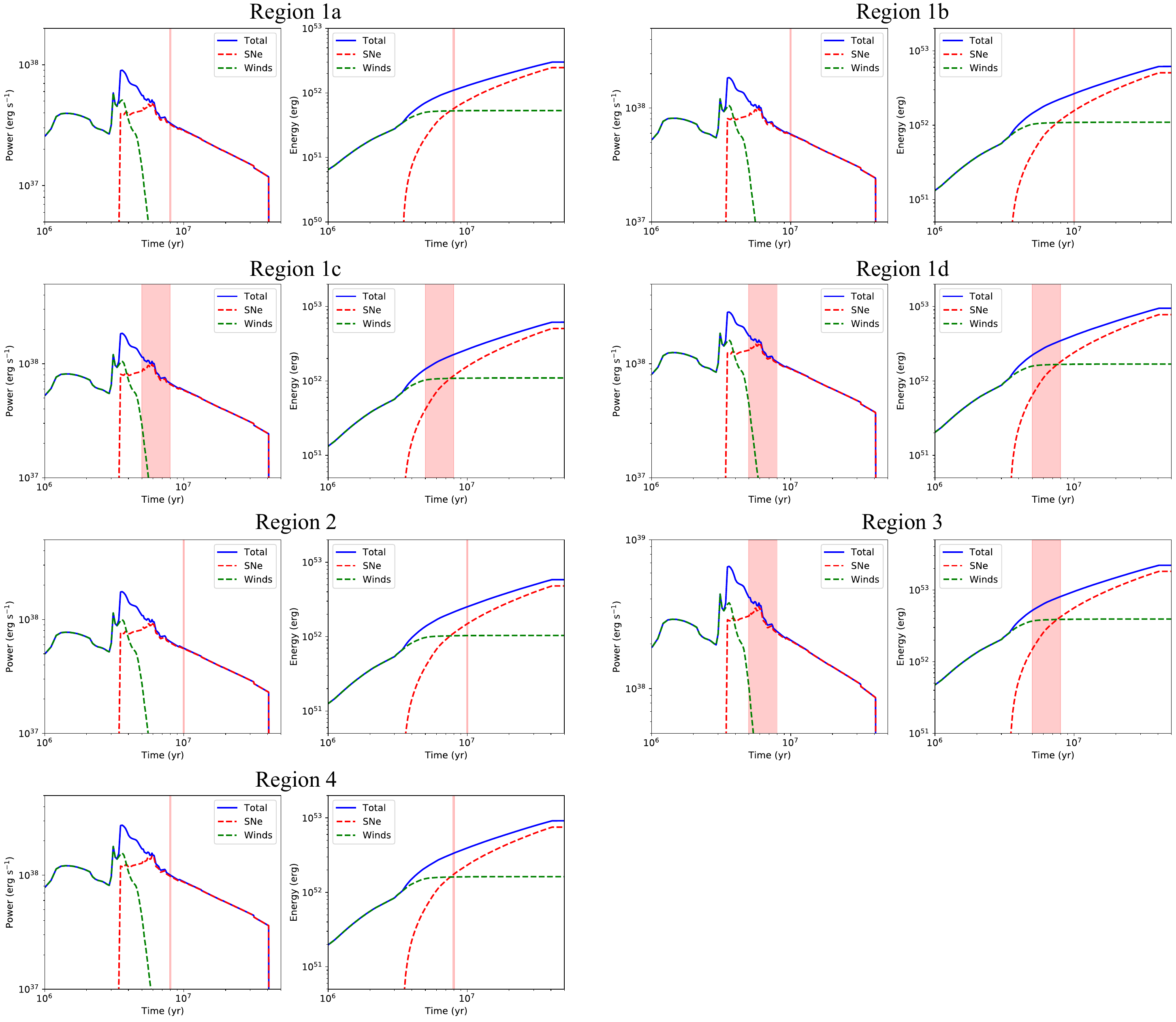}}
\caption{Wind and SN power input determined in Starburst~99 for the extended emission regions associated with MSCs. Wind power and total energy evolution are shown by the green dashed lines, SNe are shown by the red dashed lines, and the combined input by the blue solid line. The red area show the age range of the stellar populations estimated from Fig. \ref{all_cmds}.}
\label{all_sb99}
\end{center}
\end{figure*}

To assess the power input from the stellar population we use the Starburst~99 population synthesis code \citep{Leitherer1999,Vas2005,Leitherer2010}. We estimated the total stellar mass of the initial population using the numbers of stars observed in the 8-12~M$_{\sun}$ region of the colour-magnitude diagrams and assumed an initial mass function (IMF) of the form used to calculate the PARSEC evolutionary tracks. We fed this total stellar mass into Starburst~99 and assumed an instantaneous burst of star formation to produce the stellar population. The input and output are detailed for each region below. The evolution of the power output due to stellar winds and supernovae was traced over time for each region and are shown in Fig.\ref{all_sb99} along with the estimated age of the stellar populations. The stellar input properties for each region are summarised in Table~\ref{stellar_gas_properties}. 

\paragraph{Region~1:}
Since Region~1 is quite elongated along the 10~kpc ring, we sub-divided the stellar population based on aggregations of the brightest stars found in the PHAT data. This resulted in four sub-populations which we label Regions~1a--1d. We constructed a colour-magnitude diagrams  using the PHAT data (Fig.\ref{all_cmds}) and ran Starburst~99 simulations for each sub-region (Fig.\ref{all_sb99}). We found the most massive star was about 20-30~M$_{\sun}$ and ages $\lesssim10$~Myr for the MSCs. Using the Starburst~99 simulations, we estimated $\sim32-51$ SNe have occurred in these MSCs, and they are currently injecting $\sim2.5-3.9\times10^{38}$~erg~s$^{-1}$ into the region, with a total input energy over their lifetimes of $\sim6-8\times10^{52}$~erg.

\paragraph{Region~2:}
For Region~2, the PHAT colour-magnitude diagram (Fig.\ref{all_cmds}) suggested the most massive star in the MSC is about 20~M$_{\sun}$ with an age of $\sim10$~Myr for the MSC. The Starburst99 simulations (Fig.\ref{all_sb99}) provided an estimate of $\sim15$ past SNe in the region, with the MSC currently supplying $\sim5\times10^{37}$~erg~s$^{-1}$ into the region, with integrated energy input of $\sim2\times10^{52}$~erg.

\paragraph{Region~3:}
The PHAT colour-magnitude diagram (Fig.\ref{all_cmds}) suggested the most massive star in the MSC is $\sim30$~M$_{\sun}$ with an MSC age of $\sim5-8$~Myr. The Starburst99 simulations (Fig.\ref{all_sb99}) showed that $\sim14-41$ SNe may have already exploded in the region and the MSC is currently inputting $\sim2.0-4.0\times10^{38}$~erg~s$^{-1}$ into the region, with integrated energy input of $\sim5-8\times10^{52}$~erg.

\paragraph{Region~4:}
Finally, the PHAT photometry (Fig.\ref{all_cmds}) reveals the most massive star in the MSC to be $\sim20-30$~M$_{\sun}$ and the age of the MSC to be $\sim8$~Myr. Starburst99 calculated (Fig.\ref{all_sb99}) $\sim17$ SNe have occurred and the MSC is currently pumping $\sim1.0\times10^{38}$~erg~s$^{-1}$ into the region, with integrated energy input of $\sim3\times10^{52}$~erg over the MSC lifetime.

 \begin{table*}[!h]
\caption{Stellar input and hot gas properties for the X-ray enhancements in \gal.}
\begin{center}
\label{stellar_gas_properties}
{\renewcommand{\arraystretch}{1.3}
\begin{tabular}{lllllllll}
\hline
\hline
Region & M$_{\sun, \rm{max}}$ & Age &  N$_{\rm{SNe}}$  & $Pow_{\rm{SP,total}}$ & $E_{\rm{SP,total}}$ & $n$ & $P_{\rm{gas}}$ & $E_{\rm{gas}}$ \\
 &  & (Myr) &  & ($10^{37}$ erg s$^{-1}$) & ($10^{52}$ erg) & ($f^{-1/2}~10^{-2}\rm{cm}^{-3}$) & ($~f^{-1/2}~10^{4}\rm{cm}^{-3}~\rm{K}$) & ($~f^{1/2}~10^{53}$~erg) \\
\hline
1a & 20--30 & $\sim8$ & 6 & $\sim3$ & $~1$ & --  & --  & -- \\
1b & $\sim20$ & $\sim10$ & 16 & $\sim6$ & $\sim2$ & --  & --  & -- \\
1c & $\sim30$ & 5--8 & 4--11 & $6-10$ & $1-2$  & --  & --  & --  \\
1d & $\sim30$ & 5--8 & 6--18 & $10-20$ & $2-3$ & --  & --  & -- \\
1 (all) & --  & -- & -- & -- & -- & $1.5~(1.0-2.3)$ & $8.2~(4.8-14.7)$ & $4.1~(2.4-7.4)$ \\
  &   & &  &   &  &  & & \\
2 & $\sim20$ & $\sim10$ & 15 & $\sim5$ & $\sim2$ & $0.3~(0.2-0.3)$ & $4.3~(3.5-4.9)$ & $1.2~(1.0-1.3)$ \\
3 & $\sim30$ & 5--8 & 14--41 & $20-40$ & $5-8$ & $0.2~(0.1-0.4)$ & $2.0~(1.1-7.3)$ & $0.3~(0.2-1.1)$ \\
4 & $20-30$ & $\sim8$ & 17 & $\sim10$ & $\sim3$ & $0.4~(0.3-0.5)$ & $2.1~(1.8-2.5)$ & $0.9~(0.8-1.2)$\\
\hline
\end{tabular}}
\tablefoot{The values for the most massive star (M$_{\sun, \rm{max}}$) and age of the clusters were determined from the colour-magnitude diagrams in Fig.~\ref{all_cmds}. The number of SNe (N$_{\rm{SNe}}$), current power ($Pow_{\rm{SP,total}}$, and total energy $E_{\rm{SP,total}}$ supplied by the stellar population were estimated from the Starburst~99 output shown in Fig.~\ref{all_sb99} and the age estimate. The values for $n$, $P$, and E$_{\rm{gas}}$ were estimated using Eqs.~\ref{eq:n}, \ref{eq:p}, and \ref{eq:e}, respectively. The intervals for the parameters are derived from the 90\% confidence intervals of the X-ray spectral fit parameters in Table~\ref{spectral_fit_results}.
}
\end{center}
\end{table*}%

\subsubsection{Properties of the X-ray gas}
\label{gas_properties}
We can estimate the physical properties of the hot gas in the extended emission region associated with MSCs using the morphology and the results of the X-ray spectral analysis \citep[e.g.][]{Sasaki2011,Kavanagh2012}. We note that these estimates are somewhat coarse as we are limited in determining the dimensions of the emitting volumes because of the unknown depth of the emission regions in the line of sight. For Regions~2--4, which are approximately circular on the sky, we assume the volumes can be approximated as spherical. For Region~1, which is elongated along the 10~kpc ring, we assume the depth is the same as the projected width of the region (i.e. the volume is approximated as a highly elongated ellipsoid where the second and third principal axes are equal). While this is not ideal, the estimated plasma properties, though crude, can provide insight into the plasma conditions in the extended emission regions.

The gas density can be determined from the \apec\ spectral model normalisation ($K$), which is defined as:

\begin{equation}
\label{eq:n}
K = \frac{1}{10^{14}\times4 \pi D_{\rm{M~31}}^{2}}\int n_{e} n~d f V~\rm{cm}^{-5},
\end{equation}

\noindent where $D_{\rm{M~31}}$ is the distance to \gal, $n_{e}$ is the electron density, $n$ is the hydrogen density, and $f$ and $V$ are the filling parameter and volume of the X-ray emitting gas, respectively. We estimated the extent of the emission regions using the $5\sigma$ X-ray contours (see Fig.~\ref{multi_mw}) and calculated a volume assuming a distance to \gal\ of 783~kpc \citep{Conn2016}. Assuming the metallicity of young clusters in \gal\ is approximately solar, the electron-to-hydrogen ratio $n_{e}/n\approx1.21$.  Inserting the value $K$ from Table~\ref{spectral_fit_results} allowed us estimate the gas density for each region. The pressure ($P_{\rm{gas}}$) in the gas can be estimated from the gas density and the temperature determined in the spectral fits ($kT$) as

\begin{equation}
\label{eq:p}
P_{\rm{gas}}/k = 2.31n~f^{-1/2}T~\rm{cm}^{-3}~\rm{K}.
\end{equation}

\noindent Finally, the thermal energy contained in the X-ray emitting gas $E$ can be estimated from

\begin{equation}
\label{eq:e}
E_{\rm{gas}} = 3/2~P_{\rm{gas}}fV~\rm{erg}.
\end{equation}

\noindent The hot gas properties are summarised in Table~\ref{stellar_gas_properties}.

The plasma properties determined for each of our regions, bearing in mind the caveat at the beginning of this subsection, are somewhat consistent with regions around MSCs and superbubbles of the well studied Magellanic Cloud (MC) populations. The gas densities in all regions are $f^{-1/2}~10^{-(2-3)}$~cm$^{-3}$. Region~1 matches well with MC values \citep[e.g.][]{Sasaki2011,Kavanagh2012,Lopez2014,DeHorta2014}, though the density in Regions~2--4 is a factor of a few lower. This difference can most likely be explained by the difference in filling factor $f$ between the MC objects and our \gal\ regions. Studies of MC objects are well resolved on pc scales so analyses can, for the most part, be focussed on individial MSCs and superbubbles. However, in the case of our regions, we cannot resolve X-ray emission from individual objects and we are observing the combined emission from several MSCs and superbubbles and the space between these meaning that the value of $f$ is probably much lower than unity. A decrease in $f$ causes a higher value of $n$ with, e.g. $f=0.1$ giving a factor of 3 increase in $n$. If this is indeed the case, gas densities in the \gal\ regions would be in close agreement with their MC counterparts. Similarly, the derived pressures are lower than expected. The pressures in regions of active star formation are thought to be $P/k = 10^{5-6}$~cm$^{-3}$~K \citep[][and references therein]{Oey2004}, and pressures in MSCs and stellar clusters should be higher. However, we find typical values $P/k$ of $f^{-1/2}10^{4-5}$~cm$^{-3}$~K in our regions. Again, if $f$ is much lower than in the MC objects, as we suggest, this may explain the discrepancy. This should indeed be expected, since in a Milky Way like galaxy as \gal, the level of star formation implies a supernova regulated ISM. It has been shown by 3D high resolution numerical simulations \citep{Avillez2004,Avillez2005} that, e.g. in the solar neighbourhood, the volume filling factor $f_{h}$ of hot gas with $T > 10^{5.5}$~K is $f_{h} = 0.17$ (increasing to $0.19$ for a 3~$\mu$G field) for a Galactic supernova rate. This is a direct consequence of the high level of turbulence and a significant amount of gas being present in classically thermally unstable phases.

Other explanations for a decreased pressure in our regions are: 
(i) the possibility of significant blowouts into low density regions. The \ha-morphology of Region~2 exhibits a large $\sim300$~pc diameter shell which appears to have blown out, possibly out of the 10~kpc ring. Such blowouts would result in a loss in pressure in the MSCs and superbubbles (see also Sect.~\ref{lse:5}); (ii) a natural cause for lower pressure is the deviation from CIE, which is often assumed in X-ray analyses, mostly for convenience. Also NEI fit models are only a crude representation of the real physics. The reason is because the ionisation structure has to be self-consistently calculated with the thermal one. The deviation from CIE is due to the fact that most X-ray plasmas are treated in the coronal approximation, where both collisional ionisation and X-ray emission  extract energy from the plasma and are therefore cooling processes. So even if the plasma was in CIE initially, it will get out of CIE as it evolves in time \citep[for details see][]{Breitschwerdt1999,Avillez2012}. NEI will be especially enforced in regions of blowout, where adiabatic expansion of the plasma drastically reduces the kinetic energy of the electrons, while the recombination is delayed.

The determined thermal energies stored in the gas in our regions are slightly higher than expected with values of $f^{1/2}10^{52-53}$~erg when compared to typical values in the MCs ($10^{51-52}$~erg). This is not really surprising because, as mentioned above, each region contains several MSCs or superbubbles so the energy stored in a multitude of objects will naturally be higher than that in a single object as is usually studied in the MCs. The values of $E_{\rm{gas}}$ are also large when compared to the total energy input determined in the Starburst~99 simulations ($E_{\rm{SP,total}}$ in Table~\ref{stellar_gas_properties}). This, once again, is most likely due to the real value of $f$ which will act to reduce $E_{\rm{gas}}$ in the regions if $f$ is much less than unity (see Eqs.~\ref{eq:p} and \ref{eq:e}).

\subsection{Large scale emission}
\label{large_scale}
\subsubsection{Regions 5 and 6}
\label{lse:5}
The extended X-ray emission observed in Regions~5 and 6, at least in projection, appears to be located in the gap between the 5~kpc and 10~kpc rings in the northern disk (see Fig.~\ref{contour_plot}) and is not obviously associated with any MSCs (see Fig.~\ref{multiwave_images}). The X-ray emission is soft in both regions ($kT\lesssim0.2$~keV) and appears to be close to or in CIE (see Table~\ref{spectral_fit_results}). In Region~5, the emission appears brightest near the 10~kpc ring MSCs in Region~3, suggesting that the MSCs play some role in this emission. The X-ray brightness drops off suddenly midway between the 5~kpc and 10~kpc rings in the south. It is unclear if this is a real reduction in X-ray brightness or is an absorption effect resulting from our viewing angle and the 5~kpc ring blocking our line of sight closer to its northern side. Certainly, such absorption effects can be invoked to explain why there is little X-ray emission observed in the gap between the rings near Region~1 where the dust 10~kpc ring is effectively masking the gap (Fig.~\ref{multiwave_images}, top-right).

Given the close correlation between the brightest emission in Region~5 and the MSCs in Region~3, and the nearby populations of massive stars in the 5~kpc ring for Region~6, it is possible that the MSCs in the ring are supplying the hot gas observed in the gaps. From Fig~\ref{multiwave_images}, top-right and bottom-right, and Fig.~\ref{multi_mw}, bottom right, we see that the MSCs in Region~3 are projected on the inner edge of the ring, and there is an ambient density gradient increasing away from the gap. Therefore, it is possible that the MSCs have heated the hot gas which has then blown out into the lower density region and into the gap. This may also be the case for the MSCs in the 5~kpc ring supplying the hot gas in Region~6, though it is possible that the location of these MSCs on the outer edge of the ring is caused by projection. An alternative explanation for the X-ray emission observed in Region~6 is that the hot gas is a relic of an older massive stellar population, which no longer hosts the massive stars responsible for the heating. However, the cooling time for the hot plasma is $\sim10^{6-7}$~yr, so additional heating by the MSCs in the 5~kpc ring may still be required in this case.

\subsubsection{Region 7}
The extended emission in Region~7 extends from the \gal\ core out to the 5~kpc ring (excluding the emission in Region~4). This region encompasses both gaps between dust structures inside the 5~kpc ring (Fig.~\ref{multiwave_images}, top-right) and also some populations of massive stars (Fig.~\ref{multiwave_images}, bottom-right), though comparatively fewer than in the 10~kpc ring and still anti-correlated with the brightest X-ray emission. The X-ray emission is harder than in other extended regions, including those associated with MSCs and superbubbles in the 10~kpc ring, with $kT=0.94~(0.86-0.99)$~keV compared to $kT\lesssim0.7$~keV. It is unclear as to why this may be, though it is obvious that Region~7 contains a much more diverse multi-wavelength morphology and stellar population than Regions~6 and 7, so some additional heating mechanism or a longer cooling time due to lower density may be responsible. In addition, it could also be that there is some outflow component originating in Region~4, as seen in \object{NGC~5461} in \object{M~101} for example \citep{Sun2012}, projected against Region~7. The observed emission is likely an amalgam of several different emission regions and mechanisms. The low count statistics makes it difficult for us to sub-divide the region and perform a more meaningful spectral analysis using smaller spatial scales.

\subsection{Comparison with other spiral galaxies}
Early observations of nearby galaxies in X-rays with the \einstein\ Observatory and \rosat\ revealed soft diffuse X-rays from nearby spiral galaxy \citep{1992ApJS...80..531F,1995ApJ...452..627S,1995A&A...295..289E,1996ApJ...468..102C,1997MNRAS.286..626R}. 
Newer observations with \xmm\ and \chandra\ have shown that the diffuse X-ray emission in galaxy disks is found along the spiral arms and is correlated with regions with enhanced H$\alpha$, UV, or IR emission \citep{2003ApJ...588..264K,2004ApJ...610..213T}. The multi-wavelength correlation indicates that soft diffuse X-ray emission traces the regions in the disk with active star formation. The morphology and the spectrum of the diffuse emission was studied in detail in, e.g. \object{M~101} using \chandra\ and \xmm\ data by \citet{2003ApJ...588..264K} and \citet{2007MNRAS.376.1611W}, respectively.
The spectrum of the diffuse X-ray emission in the disk of spiral galaxies can be well reproduced by a thermal model with $kT = 0.20$ keV, while in some cases, an additional component with a higher temperature ($kT = 0.6 - 0.8$ keV) is required. The soft component can be explained as emission from the thermal hot phase of the ISM, while the hotter component can have contribution of late-type stars or binaries (see also Sect.\,\ref{unresolved}). We studied the nearby pair of galaxies \object{NGC~1512} and \object{NGC~1510} with \xmm. While the photon statistics were not high enough to detect diffuse emission in the outer parts of the disk, we detected significant diffuse emission in the central region of the main galaxy \object{NGC~1512} with $kT = 0.66$ keV \citep{2014A&A...566A.115D}.

In the northern disk of \gal, significant diffuse X-ray emission is detected along the major dust ring, with some additional diffuse X-ray emission filling the region inside the dust ring. Regions with bright diffuse emission seem to correlate with regions with active star formation as seen in the distribution of young massive stars in UV, \hii\ regions in H$\alpha$, or dust in IR. The temperature of the hot plasma varies from $kT$ = 0.1 -- 0.3 keV in the ring up to $kT$ = 0.6 keV in a superbubble. The region inside the major dust ring and along the inner rings appears to be filled with thermal plasma at a higher temperature ($kT$ = 0.94 keV, Region~7) even though there are no obvious bright star-forming regions. Together with the disturbed appearance of the ring which consists of narrower fainter sub-structures, the higher temperature might indicate that the ISM was not only heated by massive stars but also by some disturbances in the disk. Overall, with a few notable exceptions, the morphology and spectra of the hot gas in the northern disk of \gal\ is similar to that of other galaxy disks.

\section{Conclusion}
\label{conclusion}
We used new deep and archival \xmm\ observations of the northern disk of \gal\ to trace the hot ISM plasma in unprecedented detail and to characterise the physical properties of the X-ray emitting gas. We also used multi-wavelength surveys to put the X-ray emission in context of the multi-phase ISM. 

The brightest extended emission regions were found to be correlated with regions hosting populations of massive stars, notably in the 5~kpc and 10~kpc star-forming rings. The plasma temperatures in these regions were $\sim0.2$~keV up to $\sim0.6$~keV, which are consistent with an origin in populations of massive stars and superbubbles. The derived X-ray luminosities, densities, and pressures for the gas in each of these regions are all consistent with typical values in the literature for the well-studied Magellanic Cloud populations if the filling factor of the emission is much lower than unity. We argue that this is indeed the case for the regions studied since we don't have the spatial resolution to disentangle individual MSCs and superbubbles.

We found large extended emission filling low density gaps in the dust morphology of the northern disk, most notably between the 5~kpc and 10~kpc star-forming rings. We suggested that the hot gas in the low density regions most likely originated in the MSCs in the rings and expelled, moving down the density gradient into the gaps. 

Our deep observations of the northern disk have demonstrated that the hot ISM plasma in a galaxy like our own can be studied on relatively small spatial scales without the issues such as foreground absorption, distance uncertainties, and source confusion that affect Galactic studies. The southern disk of \gal\ offers another promising target for such a deep study. The 10~kpc ring in the southern disk hosts the large stellar association \object{NGC~206}, and is located where the ring was most likely disturbed by an encounter with the satellite galaxy \object{M~32}. Therefore, deep observations of the southern disk in X-rays would allow the study of, not only the hot ISM plasma associated with massive stellar clusters, but also the environment of the \gal, \object{M~32} collision region.

\begin{acknowledgements} 
We wish to thank the anonymous referee for their comments and suggestions which improved the paper. This research is funded by the German Bundesministerium f\"ur Wirtschaft und Technologie and the Deutsches Zentrum f\"ur Luft- und Raumfahrt (BMWi/DLR) through the grant FKZ 50 OR 1510.
M.S.\ acknowledges support by the Deutsche Forschungsgemeinschaft
through the Heisenberg professor grants SA 2131/5-1 and 12-1.
This research made use of Montage. It is funded by the National Science Foundation under Grant Number ACI-1440620, and was previously funded by the National Aeronautics and Space Administration's Earth Science Technology Office, Computation Technologies Project, under Cooperative Agreement Number NCC5-626 between NASA and the California Institute of Technology. \end{acknowledgements}

\bibliographystyle{aa}
\bibliography{refs.bib}

\begin{thebibliography}{77}
\expandafter\ifx\csname natexlab\endcsname\relax\def\natexlab#1{#1}\fi

\bibitem[{{Arnaud}(1996)}]{Arnaud1996}
{Arnaud}, K.~A. 1996, in Astronomical Society of the Pacific Conference Series,
  Vol. 101, Astronomical Data Analysis Software and Systems V, ed. G.~H.
  {Jacoby} \& J.~{Barnes}, 17

\bibitem[{{Arp}(1964)}]{Arp1964}
{Arp}, H. 1964, \apj, 139, 1045

\bibitem[{{Astropy Collaboration} {et~al.}(2013){Astropy Collaboration},
  {Robitaille}, {Tollerud}, {Greenfield}, {Droettboom}, {Bray}, {Aldcroft},
  {Davis}, {Ginsburg}, {Price-Whelan}, {Kerzendorf}, {Conley}, {Crighton},
  {Barbary}, {Muna}, {Ferguson}, {Grollier}, {Parikh}, {Nair}, {Unther},
  {Deil}, {Woillez}, {Conseil}, {Kramer}, {Turner}, {Singer}, {Fox}, {Weaver},
  {Zabalza}, {Edwards}, {Azalee Bostroem}, {Burke}, {Casey}, {Crawford},
  {Dencheva}, {Ely}, {Jenness}, {Labrie}, {Lim}, {Pierfederici}, {Pontzen},
  {Ptak}, {Refsdal}, {Servillat}, \& {Streicher}}]{Astropy2013}
{Astropy Collaboration}, {Robitaille}, T.~P., {Tollerud}, E.~J., {et~al.} 2013,
  \aap, 558, A33

\bibitem[{{Balucinska-Church} \& {McCammon}(1992)}]{Bal1992}
{Balucinska-Church}, M. \& {McCammon}, D. 1992, \apj, 400, 699

\bibitem[{{Borkowski} {et~al.}(2001){Borkowski}, {Lyerly}, \&
  {Reynolds}}]{Borkowski2001}
{Borkowski}, K.~J., {Lyerly}, W.~J., \& {Reynolds}, S.~P. 2001, \apj, 548, 820

\bibitem[{Bradley {et~al.}(2016)Bradley, Sipocz, Robitaille, Tollerud,
  Vinícius, Deil, Barbary, Günther, Cara, Droettboom, Bostroem, Bray,
  Bratholm, Pickering, Craig, Barentsen, Pascual, adonath, Greco, Kerzendorf,
  StuartLittlefair, Ferreira, D'Eugenio, \& Weaver}]{photutils2016}
Bradley, L., Sipocz, B., Robitaille, T., {et~al.} 2016, astropy/photutils: v0.3

\bibitem[{{Braun} {et~al.}(2009){Braun}, {Thilker}, {Walterbos}, \&
  {Corbelli}}]{Braun2009}
{Braun}, R., {Thilker}, D.~A., {Walterbos}, R.~A.~M., \& {Corbelli}, E. 2009,
  \apj, 695, 937

\bibitem[{{Breitschwerdt} \& {Schmutzler}(1999)}]{Breitschwerdt1999}
{Breitschwerdt}, D. \& {Schmutzler}, T. 1999, \aap, 347, 650

\bibitem[{{Bressan} {et~al.}(2012){Bressan}, {Marigo}, {Girardi}, {Salasnich},
  {Dal Cero}, {Rubele}, \& {Nanni}}]{Bressan2012}
{Bressan}, A., {Marigo}, P., {Girardi}, L., {et~al.} 2012, \mnras, 427, 127

\bibitem[{{Chabrier}(2001)}]{Chabrier2001}
{Chabrier}, G. 2001, \apj, 554, 1274

\bibitem[{{Chen} {et~al.}(1997){Chen}, {Fabian}, \& {Gendreau}}]{Chen1997}
{Chen}, L.-W., {Fabian}, A.~C., \& {Gendreau}, K.~C. 1997, \mnras, 285, 449

\bibitem[{{Chen} {et~al.}(2015){Chen}, {Bressan}, {Girardi}, {Marigo}, {Kong},
  \& {Lanza}}]{Chen2015}
{Chen}, Y., {Bressan}, A., {Girardi}, L., {et~al.} 2015, \mnras, 452, 1068

\bibitem[{{Chen} {et~al.}(2014){Chen}, {Girardi}, {Bressan}, {Marigo},
  {Barbieri}, \& {Kong}}]{Chen2014}
{Chen}, Y., {Girardi}, L., {Bressan}, A., {et~al.} 2014, \mnras, 444, 2525

\bibitem[{{Conn} {et~al.}(2016){Conn}, {McMonigal}, {Bate}, {Lewis}, {Ibata},
  {Martin}, {McConnachie}, {Ferguson}, {Irwin}, {Elahi}, {Venn}, \&
  {Mackey}}]{Conn2016}
{Conn}, A.~R., {McMonigal}, B., {Bate}, N.~F., {et~al.} 2016, \mnras, 458, 3282

\bibitem[{{Cox}(2005)}]{Cox2005}
{Cox}, D.~P. 2005, \araa, 43, 337

\bibitem[{{Cui} {et~al.}(1996){Cui}, {Sanders}, {McCammon}, {Snowden}, \&
  {Womble}}]{1996ApJ...468..102C}
{Cui}, W., {Sanders}, W.~T., {McCammon}, D., {Snowden}, S.~L., \& {Womble},
  D.~S. 1996, \apj, 468, 102

\bibitem[{{Dalcanton} {et~al.}(2012){Dalcanton}, {Williams}, {Lang}, {Lauer},
  {Kalirai}, {Seth}, {Dolphin}, {Rosenfield}, {Weisz}, {Bell}, {Bianchi},
  {Boyer}, {Caldwell}, {Dong}, {Dorman}, {Gilbert}, {Girardi}, {Gogarten},
  {Gordon}, {Guhathakurta}, {Hodge}, {Holtzman}, {Johnson}, {Larsen}, {Lewis},
  {Melbourne}, {Olsen}, {Rix}, {Rosema}, {Saha}, {Sarajedini}, {Skillman}, \&
  {Stanek}}]{Dalcanton2012}
{Dalcanton}, J.~J., {Williams}, B.~F., {Lang}, D., {et~al.} 2012, \apjs, 200,
  18

\bibitem[{{de Avillez} \& {Breitschwerdt}(2004)}]{Avillez2004}
{de Avillez}, M.~A. \& {Breitschwerdt}, D. 2004, \aap, 425, 899

\bibitem[{{de Avillez} \& {Breitschwerdt}(2005)}]{Avillez2005}
{de Avillez}, M.~A. \& {Breitschwerdt}, D. 2005, \aap, 436, 585

\bibitem[{{de Avillez} \& {Breitschwerdt}(2012)}]{Avillez2012}
{de Avillez}, M.~A. \& {Breitschwerdt}, D. 2012, \apjl, 756, L3

\bibitem[{{De Horta} {et~al.}(2014){De Horta}, {Sommer}, {Filipovi{\'c}},
  {O'Brien}, {Bozzetto}, {Collier}, {Wong}, {Crawford}, {Tothill}, {Maggi}, \&
  {Haberl}}]{DeHorta2014}
{De Horta}, A.~Y., {Sommer}, E.~R., {Filipovi{\'c}}, M.~D., {et~al.} 2014, \aj,
  147, 162

\bibitem[{{Dickey} \& {Lockman}(1990)}]{Dickey1990}
{Dickey}, J.~M. \& {Lockman}, F.~J. 1990, \araa, 28, 215

\bibitem[{{Ducci} {et~al.}(2014){Ducci}, {Kavanagh}, {Sasaki}, \&
  {Koribalski}}]{2014A&A...566A.115D}
{Ducci}, L., {Kavanagh}, P.~J., {Sasaki}, M., \& {Koribalski}, B.~S. 2014,
  \aap, 566, A115

\bibitem[{{Ehle} {et~al.}(1995){Ehle}, {Pietsch}, \&
  {Beck}}]{1995A&A...295..289E}
{Ehle}, M., {Pietsch}, W., \& {Beck}, R. 1995, \aap, 295, 289

\bibitem[{{Fabbiano} {et~al.}(1992){Fabbiano}, {Kim}, \&
  {Trinchieri}}]{1992ApJS...80..531F}
{Fabbiano}, G., {Kim}, D.~W., \& {Trinchieri}, G. 1992, \apjs, 80, 531

\bibitem[{{Fritz} {et~al.}(2012){Fritz}, {Gentile}, {Smith}, {Gear}, {Braun},
  {Duval}, {Bendo}, {Baes}, {Eales}, {Verstappen}, {Blommaert}, {Boquien},
  {Boselli}, {Clements}, {Cooray}, {Cortese}, {De Looze}, {Ford}, {Galliano},
  {Gomez}, {Gordon}, {Lebouteiller}, {O'Halloran}, {Kirk}, {Madden}, {Page},
  {Remy}, {Roussel}, {Spinoglio}, {Thilker}, {Vaccari}, {Wilson}, \&
  {Waelkens}}]{Fritz2012}
{Fritz}, J., {Gentile}, G., {Smith}, M.~W.~L., {et~al.} 2012, \aap, 546, A34

\bibitem[{{Galvin} \& {Filipovic}(2014)}]{Galvin2014}
{Galvin}, T.~J. \& {Filipovic}, M.~D. 2014, Serbian Astronomical Journal, 189,
  15

\bibitem[{{Galvin} {et~al.}(2012){Galvin}, {Filipovic}, {Crawford}, {Tothill},
  {Wong}, \& {De Horta}}]{Galvin2012}
{Galvin}, T.~J., {Filipovic}, M.~D., {Crawford}, E.~J., {et~al.} 2012, Serbian
  Astronomical Journal, 184, 41

\bibitem[{{Gordon} {et~al.}(2006){Gordon}, {Bailin}, {Engelbracht}, {Rieke},
  {Misselt}, {Latter}, {Young}, {Ashby}, {Barmby}, {Gibson}, {Hines}, {Hinz},
  {Krause}, {Levine}, {Marleau}, {Noriega-Crespo}, {Stolovy}, {Thilker}, \&
  {Werner}}]{Gordon2006}
{Gordon}, K.~D., {Bailin}, J., {Engelbracht}, C.~W., {et~al.} 2006, \apjl, 638,
  L87

\bibitem[{{Griffin} {et~al.}(2010){Griffin}, {Abergel}, {Abreu}, {Ade},
  {Andr{\'e}}, {Augueres}, {Babbedge}, {Bae}, {Baillie}, {Baluteau}, {Barlow},
  {Bendo}, {Benielli}, {Bock}, {Bonhomme}, {Brisbin}, {Brockley-Blatt},
  {Caldwell}, {Cara}, {Castro-Rodriguez}, {Cerulli}, {Chanial}, {Chen},
  {Clark}, {Clements}, {Clerc}, {Coker}, {Communal}, {Conversi}, {Cox},
  {Crumb}, {Cunningham}, {Daly}, {Davis}, {de Antoni}, {Delderfield}, {Devin},
  {di Giorgio}, {Didschuns}, {Dohlen}, {Donati}, {Dowell}, {Dowell}, {Duband},
  {Dumaye}, {Emery}, {Ferlet}, {Ferrand}, {Fontignie}, {Fox}, {Franceschini},
  {Frerking}, {Fulton}, {Garcia}, {Gastaud}, {Gear}, {Glenn}, {Goizel},
  {Griffin}, {Grundy}, {Guest}, {Guillemet}, {Hargrave}, {Harwit}, {Hastings},
  {Hatziminaoglou}, {Herman}, {Hinde}, {Hristov}, {Huang}, {Imhof}, {Isaak},
  {Israelsson}, {Ivison}, {Jennings}, {Kiernan}, {King}, {Lange}, {Latter},
  {Laurent}, {Laurent}, {Leeks}, {Lellouch}, {Levenson}, {Li}, {Li},
  {Lilienthal}, {Lim}, {Liu}, {Lu}, {Madden}, {Mainetti}, {Marliani}, {McKay},
  {Mercier}, {Molinari}, {Morris}, {Moseley}, {Mulder}, {Mur}, {Naylor},
  {Nguyen}, {O'Halloran}, {Oliver}, {Olofsson}, {Olofsson}, {Orfei}, {Page},
  {Pain}, {Panuzzo}, {Papageorgiou}, {Parks}, {Parr-Burman}, {Pearce},
  {Pearson}, {P{\'e}rez-Fournon}, {Pinsard}, {Pisano}, {Podosek}, {Pohlen},
  {Polehampton}, {Pouliquen}, {Rigopoulou}, {Rizzo}, {Roseboom}, {Roussel},
  {Rowan-Robinson}, {Rownd}, {Saraceno}, {Sauvage}, {Savage}, {Savini},
  {Sawyer}, {Scharmberg}, {Schmitt}, {Schneider}, {Schulz}, {Schwartz},
  {Shafer}, {Shupe}, {Sibthorpe}, {Sidher}, {Smith}, {Smith}, {Smith},
  {Spencer}, {Stobie}, {Sudiwala}, {Sukhatme}, {Surace}, {Stevens}, {Swinyard},
  {Trichas}, {Tourette}, {Triou}, {Tseng}, {Tucker}, {Turner}, {Vaccari},
  {Valtchanov}, {Vigroux}, {Virique}, {Voellmer}, {Walker}, {Ward}, {Waskett},
  {Weilert}, {Wesson}, {White}, {Whitehouse}, {Wilson}, {Winter}, {Woodcraft},
  {Wright}, {Xu}, {Zavagno}, {Zemcov}, {Zhang}, \& {Zonca}}]{Griffin2010}
{Griffin}, M.~J., {Abergel}, A., {Abreu}, A., {et~al.} 2010, \aap, 518, L3

\bibitem[{{Habing} {et~al.}(1984){Habing}, {Miley}, {Young}, {Baud}, {Boggess},
  {Clegg}, {de Jong}, {Harris}, {Raimond}, {Rowan-Robinson}, \&
  {Soifer}}]{Habing1984}
{Habing}, H.~J., {Miley}, G., {Young}, E., {et~al.} 1984, \apjl, 278, L59

\bibitem[{{Jansen} {et~al.}(2001){Jansen}, {Lumb}, {Altieri}, {Clavel}, {Ehle},
  {Erd}, {Gabriel}, {Guainazzi}, {Gondoin}, {Much}, {Munoz}, {Santos},
  {Schartel}, {Texier}, \& {Vacanti}}]{Jansen2001}
{Jansen}, F., {Lumb}, D., {Altieri}, B., {et~al.} 2001, \aap, 365, L1

\bibitem[{{Jarrett} {et~al.}(2003){Jarrett}, {Chester}, {Cutri}, {Schneider},
  \& {Huchra}}]{Jarrett2003}
{Jarrett}, T.~H., {Chester}, T., {Cutri}, R., {Schneider}, S.~E., \& {Huchra},
  J.~P. 2003, \aj, 125, 525

\bibitem[{{Johnson} {et~al.}(2016){Johnson}, {Seth}, {Dalcanton}, {Beerman},
  {Fouesneau}, {Lewis}, {Weisz}, {Williams}, {Bell}, {Dolphin}, {Larsen},
  {Sandstrom}, \& {Skillman}}]{Johnson2016}
{Johnson}, L.~C., {Seth}, A.~C., {Dalcanton}, J.~J., {et~al.} 2016, \apj, 827,
  33

\bibitem[{{Johnson} {et~al.}(2012){Johnson}, {Seth}, {Dalcanton}, {Caldwell},
  {Fouesneau}, {Gouliermis}, {Hodge}, {Larsen}, {Olsen}, {San Roman},
  {Sarajedini}, {Weisz}, {Williams}, {Beerman}, {Bianchi}, {Dolphin},
  {Girardi}, {Guhathakurta}, {Kalirai}, {Lang}, {Monachesi}, {Nanda}, {Rix}, \&
  {Skillman}}]{Johnson2012}
{Johnson}, L.~C., {Seth}, A.~C., {Dalcanton}, J.~J., {et~al.} 2012, \apj, 752,
  95

\bibitem[{{Johnson} {et~al.}(2015){Johnson}, {Seth}, {Dalcanton}, {Wallace},
  {Simpson}, {Lintott}, {Kapadia}, {Skillman}, {Caldwell}, {Fouesneau},
  {Weisz}, {Williams}, {Beerman}, {Gouliermis}, \& {Sarajedini}}]{Johnson2015}
{Johnson}, L.~C., {Seth}, A.~C., {Dalcanton}, J.~J., {et~al.} 2015, \apj, 802,
  127

\bibitem[{{Kavanagh} {et~al.}(2012){Kavanagh}, {Sasaki}, \&
  {Points}}]{Kavanagh2012}
{Kavanagh}, P.~J., {Sasaki}, M., \& {Points}, S.~D. 2012, \aap, 547, A19

\bibitem[{{Kuntz} \& {Snowden}(2008)}]{Kuntz2008}
{Kuntz}, K.~D. \& {Snowden}, S.~L. 2008, \aap, 478, 575

\bibitem[{{Kuntz} \& {Snowden}(2010)}]{Kuntz2010}
{Kuntz}, K.~D. \& {Snowden}, S.~L. 2010, \apjs, 188, 46

\bibitem[{{Kuntz} {et~al.}(2003){Kuntz}, {Snowden}, {Pence}, \&
  {Mukai}}]{2003ApJ...588..264K}
{Kuntz}, K.~D., {Snowden}, S.~L., {Pence}, W.~D., \& {Mukai}, K. 2003, \apj,
  588, 264

\bibitem[{{Leitherer} {et~al.}(2010){Leitherer}, {Ortiz Ot{\'a}lvaro},
  {Bresolin}, {Kudritzki}, {Lo Faro}, {Pauldrach}, {Pettini}, \&
  {Rix}}]{Leitherer2010}
{Leitherer}, C., {Ortiz Ot{\'a}lvaro}, P.~A., {Bresolin}, F., {et~al.} 2010,
  \apjs, 189, 309

\bibitem[{{Leitherer} {et~al.}(1999){Leitherer}, {Schaerer}, {Goldader},
  {Delgado}, {Robert}, {Kune}, {de Mello}, {Devost}, \&
  {Heckman}}]{Leitherer1999}
{Leitherer}, C., {Schaerer}, D., {Goldader}, J.~D., {et~al.} 1999, \apjs, 123,
  3

\bibitem[{{Lewis} {et~al.}(2015){Lewis}, {Dolphin}, {Dalcanton}, {Weisz},
  {Williams}, {Bell}, {Seth}, {Simones}, {Skillman}, {Choi}, {Fouesneau},
  {Guhathakurta}, {Johnson}, {Kalirai}, {Leroy}, {Monachesi}, {Rix}, \&
  {Schruba}}]{Lewis2015}
{Lewis}, A.~R., {Dolphin}, A.~E., {Dalcanton}, J.~J., {et~al.} 2015, \apj, 805,
  183

\bibitem[{{Li} \& {Wang}(2007)}]{Li2007}
{Li}, Z. \& {Wang}, Q.~D. 2007, \apjl, 668, L39

\bibitem[{{Lopez} {et~al.}(2014){Lopez}, {Krumholz}, {Bolatto}, {Prochaska},
  {Ramirez-Ruiz}, \& {Castro}}]{Lopez2014}
{Lopez}, L.~A., {Krumholz}, M.~R., {Bolatto}, A.~D., {et~al.} 2014, \apj, 795,
  121

\bibitem[{{Maggi} {et~al.}(2016){Maggi}, {Haberl}, {Kavanagh}, {Sasaki},
  {Bozzetto}, {Filipovi{\'c}}, {Vasilopoulos}, {Pietsch}, {Points}, {Chu},
  {Dickel}, {Ehle}, {Williams}, \& {Greiner}}]{Maggi2016}
{Maggi}, P., {Haberl}, F., {Kavanagh}, P.~J., {et~al.} 2016, \aap, 585, A162

\bibitem[{{Massey} {et~al.}(2006){Massey}, {Olsen}, {Hodge}, {Strong},
  {Jacoby}, {Schlingman}, \& {Smith}}]{Massey2006}
{Massey}, P., {Olsen}, K.~A.~G., {Hodge}, P.~W., {et~al.} 2006, \aj, 131, 2478

\bibitem[{{Oey} \& {Garc{\'{\i}}a-Segura}(2004)}]{Oey2004}
{Oey}, M.~S. \& {Garc{\'{\i}}a-Segura}, G. 2004, \apj, 613, 302

\bibitem[{{Pilbratt} {et~al.}(2010){Pilbratt}, {Riedinger}, {Passvogel},
  {Crone}, {Doyle}, {Gageur}, {Heras}, {Jewell}, {Metcalfe}, {Ott}, \&
  {Schmidt}}]{Pilbratt2010}
{Pilbratt}, G.~L., {Riedinger}, J.~R., {Passvogel}, T., {et~al.} 2010, \aap,
  518, L1

\bibitem[{{Read} {et~al.}(1997){Read}, {Ponman}, \&
  {Strickland}}]{1997MNRAS.286..626R}
{Read}, A.~M., {Ponman}, T.~J., \& {Strickland}, D.~K. 1997, \mnras, 286, 626

\bibitem[{{Revnivtsev} {et~al.}(2007){Revnivtsev}, {Churazov}, {Sazonov},
  {Forman}, \& {Jones}}]{Revnivtsev2007}
{Revnivtsev}, M., {Churazov}, E., {Sazonov}, S., {Forman}, W., \& {Jones}, C.
  2007, \aap, 473, 783

\bibitem[{{Revnivtsev} {et~al.}(2008){Revnivtsev}, {Churazov}, {Sazonov},
  {Forman}, \& {Jones}}]{Revnivtsev2008}
{Revnivtsev}, M., {Churazov}, E., {Sazonov}, S., {Forman}, W., \& {Jones}, C.
  2008, \aap, 490, 37

\bibitem[{{Rieke} {et~al.}(2004){Rieke}, {Young}, {Engelbracht}, {Kelly},
  {Low}, {Haller}, {Beeman}, {Gordon}, {Stansberry}, {Misselt}, {Cadien},
  {Morrison}, {Rivlis}, {Latter}, {Noriega-Crespo}, {Padgett}, {Stapelfeldt},
  {Hines}, {Egami}, {Muzerolle}, {Alonso-Herrero}, {Blaylock}, {Dole}, {Hinz},
  {Le Floc'h}, {Papovich}, {P{\'e}rez-Gonz{\'a}lez}, {Smith}, {Su}, {Bennett},
  {Frayer}, {Henderson}, {Lu}, {Masci}, {Pesenson}, {Rebull}, {Rho}, {Keene},
  {Stolovy}, {Wachter}, {Wheaton}, {Werner}, \& {Richards}}]{Rieke2004}
{Rieke}, G.~H., {Young}, E.~T., {Engelbracht}, C.~W., {et~al.} 2004, \apjs,
  154, 25

\bibitem[{{Salpeter}(1955)}]{Salpeter1955}
{Salpeter}, E.~E. 1955, \apj, 121, 161

\bibitem[{{Sasaki} {et~al.}(2011){Sasaki}, {Breitschwerdt}, {Baumgartner}, \&
  {Haberl}}]{Sasaki2011}
{Sasaki}, M., {Breitschwerdt}, D., {Baumgartner}, V., \& {Haberl}, F. 2011,
  \aap, 528, A136

\bibitem[{{Sasaki} {et~al.}(2018){Sasaki}, {Haberl}, {Henze}, {Saeedi},
  {Williams}, {Plucinsky}, {Hatzidimitriou}, {Karampelas}, {Sokolovsky},
  {Breitschwerdt}, {de Avillez}, {Filipovi{\'c}}, {Galvin}, {Kavanagh}, \&
  {Long}}]{2018A&A...620A..28S}
{Sasaki}, M., {Haberl}, F., {Henze}, M., {et~al.} 2018, \aap, 620, A28

\bibitem[{{Sazonov} {et~al.}(2006){Sazonov}, {Revnivtsev}, {Gilfanov},
  {Churazov}, \& {Sunyaev}}]{Sazonov2006}
{Sazonov}, S., {Revnivtsev}, M., {Gilfanov}, M., {Churazov}, E., \& {Sunyaev},
  R. 2006, \aap, 450, 117

\bibitem[{{Smith} {et~al.}(2012){Smith}, {Eales}, {Gomez}, {Roman-Duval},
  {Fritz}, {Braun}, {Baes}, {Bendo}, {Blommaert}, {Boquien}, {Boselli},
  {Clements}, {Cooray}, {Cortese}, {De Looze}, {Ford}, {Gear}, {Gentile},
  {Gordon}, {Kirk}, {Lebouteiller}, {Madden}, {Mentuch}, {O'Halloran}, {Page},
  {Schulz}, {Spinoglio}, {Verstappen}, {Wilson}, \& {Thilker}}]{Smith2012}
{Smith}, M.~W.~L., {Eales}, S.~A., {Gomez}, H.~L., {et~al.} 2012, \apj, 756, 40

\bibitem[{{Smith} {et~al.}(2001){Smith}, {Brickhouse}, {Liedahl}, \&
  {Raymond}}]{Smith2001}
{Smith}, R.~K., {Brickhouse}, N.~S., {Liedahl}, D.~A., \& {Raymond}, J.~C.
  2001, \apjl, 556, L91

\bibitem[{{Snowden} {et~al.}(2004){Snowden}, {Collier}, \&
  {Kuntz}}]{Snowden2004}
{Snowden}, S.~L., {Collier}, M.~R., \& {Kuntz}, K.~D. 2004, \apj, 610, 1182

\bibitem[{{Snowden} {et~al.}(2008){Snowden}, {Mushotzky}, {Kuntz}, \&
  {Davis}}]{Snowden2008}
{Snowden}, S.~L., {Mushotzky}, R.~F., {Kuntz}, K.~D., \& {Davis}, D.~S. 2008,
  \aap, 478, 615

\bibitem[{{Snowden} \& {Pietsch}(1995)}]{1995ApJ...452..627S}
{Snowden}, S.~L. \& {Pietsch}, W. 1995, \apj, 452, 627

\bibitem[{{Stetson}(1987)}]{Stetson1987}
{Stetson}, P.~B. 1987, \pasp, 99, 191

\bibitem[{{Stiele} {et~al.}(2011){Stiele}, {Pietsch}, {Haberl},
  {Hatzidimitriou}, {Barnard}, {Williams}, {Kong}, \& {Kolb}}]{Stiele2011}
{Stiele}, H., {Pietsch}, W., {Haberl}, F., {et~al.} 2011, \aap, 534, A55

\bibitem[{{Str{\"u}der} {et~al.}(2001){Str{\"u}der}, {Briel}, {Dennerl},
  {Hartmann}, {Kendziorra}, {Meidinger}, {Pfeffermann}, {Reppin}, {Aschenbach},
  {Bornemann}, {Br{\"a}uninger}, {Burkert}, {Elender}, {Freyberg}, {Haberl},
  {Hartner}, {Heuschmann}, {Hippmann}, {Kastelic}, {Kemmer}, {Kettenring},
  {Kink}, {Krause}, {M{\"u}ller}, {Oppitz}, {Pietsch}, {Popp}, {Predehl},
  {Read}, {Stephan}, {St{\"o}tter}, {Tr{\"u}mper}, {Holl}, {Kemmer}, {Soltau},
  {St{\"o}tter}, {Weber}, {Weichert}, {von Zanthier}, {Carathanassis}, {Lutz},
  {Richter}, {Solc}, {B{\"o}ttcher}, {Kuster}, {Staubert}, {Abbey}, {Holland},
  {Turner}, {Balasini}, {Bignami}, {La Palombara}, {Villa}, {Buttler},
  {Gianini}, {Lain{\'e}}, {Lumb}, \& {Dhez}}]{Struder2001}
{Str{\"u}der}, L., {Briel}, U., {Dennerl}, K., {et~al.} 2001, \aap, 365, L18

\bibitem[{{Sturm}(2012)}]{SturmPhD}
{Sturm}, R.~K.~N. 2012, PhD thesis, Fakult{\"a}t f{\"u}r Physik, Technische
  Universit{\"a}t M{\"u}nchen, Germany

\bibitem[{{Sun} {et~al.}(2012){Sun}, {Chen}, {Feng}, {Chu}, {Chen}, {Wang}, \&
  {Li}}]{Sun2012}
{Sun}, W., {Chen}, Y., {Feng}, L., {et~al.} 2012, \apj, 760, 61

\bibitem[{{Tang} {et~al.}(2014){Tang}, {Bressan}, {Rosenfield}, {Slemer},
  {Marigo}, {Girardi}, \& {Bianchi}}]{Tang2014}
{Tang}, J., {Bressan}, A., {Rosenfield}, P., {et~al.} 2014, \mnras, 445, 4287

\bibitem[{{Thilker} {et~al.}(2005){Thilker}, {Hoopes}, {Bianchi}, {Boissier},
  {Rich}, {Seibert}, {Friedman}, {Rey}, {Buat}, {Barlow}, {Byun}, {Donas},
  {Forster}, {Heckman}, {Jelinsky}, {Lee}, {Madore}, {Malina}, {Martin},
  {Milliard}, {Morrissey}, {Neff}, {Schiminovich}, {Siegmund}, {Small},
  {Szalay}, {Welsh}, \& {Wyder}}]{Thilker2005}
{Thilker}, D.~A., {Hoopes}, C.~G., {Bianchi}, L., {et~al.} 2005, \apjl, 619,
  L67

\bibitem[{{Turner} {et~al.}(2001){Turner}, {Abbey}, {Arnaud}, {Balasini},
  {Barbera}, {Belsole}, {Bennie}, {Bernard}, {Bignami}, {Boer}, {Briel},
  {Butler}, {Cara}, {Chabaud}, {Cole}, {Collura}, {Conte}, {Cros}, {Denby},
  {Dhez}, {Di Coco}, {Dowson}, {Ferrando}, {Ghizzardi}, {Gianotti}, {Goodall},
  {Gretton}, {Griffiths}, {Hainaut}, {Hochedez}, {Holland}, {Jourdain},
  {Kendziorra}, {Lagostina}, {Laine}, {La Palombara}, {Lortholary}, {Lumb},
  {Marty}, {Molendi}, {Pigot}, {Poindron}, {Pounds}, {Reeves}, {Reppin},
  {Rothenflug}, {Salvetat}, {Sauvageot}, {Schmitt}, {Sembay}, {Short},
  {Spragg}, {Stephen}, {Str{\"u}der}, {Tiengo}, {Trifoglio}, {Tr{\"u}mper},
  {Vercellone}, {Vigroux}, {Villa}, {Ward}, {Whitehead}, \&
  {Zonca}}]{Turner2001}
{Turner}, M.~J.~L., {Abbey}, A., {Arnaud}, M., {et~al.} 2001, \aap, 365, L27

\bibitem[{{Tyler} {et~al.}(2004){Tyler}, {Quillen}, {LaPage}, \&
  {Rieke}}]{2004ApJ...610..213T}
{Tyler}, K., {Quillen}, A.~C., {LaPage}, A., \& {Rieke}, G.~H. 2004, \apj, 610,
  213

\bibitem[{{V{\'a}zquez} \& {Leitherer}(2005)}]{Vas2005}
{V{\'a}zquez}, G.~A. \& {Leitherer}, C. 2005, \apj, 621, 695

\bibitem[{{Warwick} {et~al.}(2007){Warwick}, {Jenkins}, {Read}, {Roberts}, \&
  {Owen}}]{2007MNRAS.376.1611W}
{Warwick}, R.~S., {Jenkins}, L.~P., {Read}, A.~M., {Roberts}, T.~P., \& {Owen},
  R.~A. 2007, \mnras, 376, 1611

\bibitem[{{Werner} {et~al.}(2004){Werner}, {Roellig}, {Low}, {Rieke}, {Rieke},
  {Hoffmann}, {Young}, {Houck}, {Brandl}, {Fazio}, {Hora}, {Gehrz}, {Helou},
  {Soifer}, {Stauffer}, {Keene}, {Eisenhardt}, {Gallagher}, {Gautier}, {Irace},
  {Lawrence}, {Simmons}, {Van Cleve}, {Jura}, {Wright}, \&
  {Cruikshank}}]{Werner2004}
{Werner}, M.~W., {Roellig}, T.~L., {Low}, F.~J., {et~al.} 2004, \apjs, 154, 1

\bibitem[{{Williams}(2003)}]{Williams2003}
{Williams}, B.~F. 2003, \aj, 126, 1312

\bibitem[{{Williams} {et~al.}(2018){Williams}, {Lazzarini}, {Plucinsky},
  {Sasaki}, {Antoniou}, {Vulic}, {Eracleous}, {Long}, {Binder}, {Dalcanton},
  {Lewis}, \& {Weisz}}]{2018ApJS..239...13W}
{Williams}, B.~F., {Lazzarini}, M., {Plucinsky}, P.~P., {et~al.} 2018, \apjs,
  239, 13

\bibitem[{{Wilms} {et~al.}(2000){Wilms}, {Allen}, \& {McCray}}]{Wilms2000}
{Wilms}, J., {Allen}, A., \& {McCray}, R. 2000, \apj, 542, 914

\end{thebibliography}

\onecolumn
\clearpage
\begin{appendix}
\section{Table of observations}
\begin{small}
\begin{longtable}{llllllccc}
\caption{\label{obscat}\xmm\ observations of \gal. Exposure times are flare-filtered.}\\
\hline
Obs. ID	 & Object	 & PI	 & Obs. date	 & RA 	 & Dec	 & \multicolumn{3}{c}{EPIC exposure times (s)} \\ 
 & & & &  &  & pn	 & MOS1	 & MOS2 \\ 
\hline
\hline
\endfirsthead
\caption{continued.}\\
\hline
Obs. ID	 & Object	 & PI	 & Obs. date	 & RA	 & Dec 	 & \multicolumn{3}{c}{EPIC exposure times (s)} \\ 
 & & & &  &  & pn 	 & MOS1 	 & MOS2  \\ 
\hline
\hline
\endhead
0109270101	 	&	 M31Core	 	&	 K. Mason	 	&	 2001-06-29	 	&	 00:42:43.0	 	&	 41:15:46	 	&	 23220	  	&	 27357	  	&	 28677	 \\ 
0109270301	 	&	 M31North2	 	&	 K. Mason	 	&	 2002-01-26	 	&	 00:45:20.0	 	&	 41:56:08	 	&	 24300	  	&	 25774	  	&	 25696	 \\ 
0109270401	 	&	 M31North3	 	&	 K. Mason	 	&	 2002-06-29	 	&	 00:46:38.0	 	&	 42:16:20	 	&	 36174	  	&	 46967	  	&	 46784	 \\ 
0109270701	 	&	 M31North1	 	&	 K. Mason	 	&	 2002-01-05	 	&	 00:44:01.0	 	&	 41:35:56	 	&	 52320	  	&	 55422	  	&	 55649	 \\
0112570101	 	&	 M31 Core	 	&	 M. Watson	 	&	 2002-01-06	 	&	 00:42:43.0	 	&	 41:15:46	 	&	 50391	  	&	 55651	  	&	 56092	 \\ 
0112570201	 	&	 M31 South 1	 	&	 M. Watson	 	&	 2002-01-12	 	&	 00:41:25.0	 	&	 40:55:35	 	&	 49282	  	&	 53499	  	&	 53426	 \\ 
0112570301	 	&	 M31 South 2	 	&	 M. Watson	 	&	 2002-01-24	 	&	 00:40:06.0	 	&	 40:35:24	 	&	 20820	  	&	 25628	  	&	 26580	 \\ 
0112570401	 	&	 M31 Core	 	&	 M. Watson	 	&	 2000-06-25	 	&	 00:42:43.0	 	&	 41:15:46	 	&	 25260	  	&	 28977	  	&	 29337	 \\ 
0202230401	 	&	 RX J0042.6+4115	 	&	 R. Barnard	 	&	 2004-07-19	 	&	 00:42:38.6	 	&	 41:16:03	 	&	 --	  	&	 19806	  	&	 19871	 \\ 
0402560201	 	&	 M 31 SS1	 	&	 W. Pietsch	 	&	 2006-06-30	 	&	 00:43:28.8	 	&	 40:55:12	 	&	 --	  	&	 20280	  	&	 22320	 \\ 
0402560401	 	&	 M 31 SS2	 	&	 W. Pietsch	 	&	 2006-07-08	 	&	 00:42:16.8	 	&	 40:37:11	 	&	 --	  	&	 33405	  	&	 37490	 \\ 
0402560501	 	&	 M 31 SN2	 	&	 W. Pietsch	 	&	 2006-07-20	 	&	 00:39:40.8	 	&	 40:58:47	 	&	 29091	  	&	 50926	  	&	 54230	 \\ 
0402560701	 	&	 M 31 SN3	 	&	 W. Pietsch	 	&	 2006-07-23	 	&	 00:39:02.4	 	&	 40:37:48	 	&	 17160	  	&	 26337	  	&	 28137	 \\ 
0402560801	 	&	 M 31 S2	 	&	 W. Pietsch	 	&	 2006-12-25	 	&	 00:40:06.0	 	&	 40:35:24	 	&	 41807	  	&	 47559	  	&	 47627	 \\ 
0402560901	 	&	 M 31 NN1	 	&	 W. Pietsch	 	&	 2006-12-26	 	&	 00:41:52.8	 	&	 41:36:35	 	&	 35070	  	&	 43812	  	&	 45051	 \\ 
0402561001	 	&	 M 31 NS1	 	&	 W. Pietsch	 	&	 2006-12-30	 	&	 00:44:38.4	 	&	 41:12:00	 	&	 41419	  	&	 49663	  	&	 51228	 \\ 
0402561101	 	&	 M 31 NN2	 	&	 W. Pietsch	 	&	 2007-01-01	 	&	 00:43:09.6	 	&	 41:55:12	 	&	 35880	  	&	 45492	  	&	 46080	 \\ 
0402561201	 	&	 M 31 NS2	 	&	 W. Pietsch	 	&	 2007-01-02	 	&	 00:45:43.2	 	&	 41:31:48	 	&	 34620	  	&	 39897	  	&	 39177	 \\ 
0402561301	 	&	 M 31 NN3	 	&	 W. Pietsch	 	&	 2007-01-03	 	&	 00:44:45.6	 	&	 42:09:35	 	&	 27810	  	&	 34113	  	&	 34126	 \\ 
0402561401	 	&	 M 31 NS3	 	&	 W. Pietsch	 	&	 2007-01-04	 	&	 00:46:38.4	 	&	 41:53:59	 	&	 42234	  	&	 44997	  	&	 44397	 \\ 
0402561501	 	&	 M 31 N2	 	&	 W. Pietsch	 	&	 2007-01-05	 	&	 00:45:20.0	 	&	 41:56:08	 	&	 35640	  	&	 41604	  	&	 41382	 \\ 
0405320701	 	&	 M31	 	&	 W. Pietsch	 	&	 2006-12-31	 	&	 00:42:44.3	 	&	 41:16:09	 	&	 --	  	&	 15060	  	&	 --	 \\ 
0505720201	 	&	 M31	 	&	 W. Pietsch	 	&	 2007-12-29	 	&	 00:42:44.3	 	&	 41:16:09	 	&	 25020	  	&	 26097	  	&	 26337	 \\ 
0505720301	 	&	 M31	 	&	 W. Pietsch	 	&	 2008-01-08	 	&	 00:42:44.3	 	&	 41:16:09	 	&	 21938	  	&	 25716	  	&	 25538	 \\ 
0505720601	 	&	 M31	 	&	 W. Pietsch	 	&	 2008-02-07	 	&	 00:42:44.3	 	&	 41:16:09	 	&	 --	  	&	 20354	  	&	 19639	 \\ 
0505760101	 	&	 M 31 S3	 	&	 W. Pietsch	 	&	 2007-07-24	 	&	 00:38:52.8	 	&	 40:15:00	 	&	 16738	  	&	 25018	  	&	 30337	 \\ 
0505760201	 	&	 M 31 SS1	 	&	 W. Pietsch	 	&	 2007-07-22	 	&	 00:43:28.8	 	&	 40:55:12	 	&	 23820	  	&	 37711	  	&	 44851	 \\ 
0505760301	 	&	 M 31 SS2	 	&	 W. Pietsch	 	&	 2007-12-29	 	&	 00:42:16.8	 	&	 40:37:11	 	&	 38642	  	&	 40865	  	&	 42743	 \\ 
0505760401	 	&	 M 31 SS3	 	&	 W. Pietsch	 	&	 2007-12-25	 	&	 00:40:45.6	 	&	 40:21:00	 	&	 24945	  	&	 27536	  	&	 27237	 \\ 
0505760501	 	&	 M 31 SN3	 	&	 W. Pietsch	 	&	 2007-12-31	 	&	 00:39:02.4	 	&	 40:37:48	 	&	 24877	  	&	 33096	  	&	 33821	 \\ 
0511380101	 	&	 M 31 S3	 	&	 W. Pietsch	 	&	 2008-01-02	 	&	 00:38:52.8	 	&	 40:15:00	 	&	 40570	  	&	 44115	  	&	 44600	 \\ 
0511380201	 	&	 M 31 SS1	 	&	 W. Pietsch	 	&	 2008-01-05	 	&	 00:43:28.8	 	&	 40:55:12	 	&	 --	  	&	 16612	  	&	 16310	 \\ 
0511380301	 	&	 M 31 SN2	 	&	 W. Pietsch	 	&	 2008-01-06	 	&	 00:39:40.8	 	&	 40:58:47	 	&	 27048	  	&	 31606	  	&	 31969	 \\ 
0511380601	 	&	 M 31 SS1	 	&	 W. Pietsch	 	&	 2008-02-09	 	&	 00:43:28.8	 	&	 40:55:12	 	&	 --	  	&	 17697	  	&	 17697	 \\ 
0551690201	 	&	 M31	 	&	 W. Pietsch	 	&	 2008-12-30	 	&	 00:42:44.3	 	&	 41:16:09	 	&	 --	  	&	 20894	  	&	 20824	 \\ 
0551690301	 	&	 M31	 	&	 W. Pietsch	 	&	 2009-01-09	 	&	 00:42:44.3	 	&	 41:16:09	 	&	 17330	  	&	 19864	  	&	 19813	 \\ 
0551690501	 	&	 M31	 	&	 W. Pietsch	 	&	 2009-01-27	 	&	 00:42:44.3	 	&	 41:16:09	 	&	 --	  	&	 17177	  	&	 18489	 \\ 
0600660201	 	&	 M31	 	&	 W. Pietsch	 	&	 2009-12-28	 	&	 00:42:44.3	 	&	 41:16:09	 	&	 15894	  	&	 17936	  	&	 17157	 \\ 
0600660301	 	&	 M31	 	&	 W. Pietsch	 	&	 2010-01-07	 	&	 00:42:44.3	 	&	 41:16:09	 	&	 --	  	&	 16355	  	&	 16240	 \\ 
0600660401	 	&	 M31	 	&	 W. Pietsch	 	&	 2010-01-15	 	&	 00:42:44.3	 	&	 41:16:09	 	&	 --	  	&	 --	  	&	 15657	 \\ 
0600660501	 	&	 M31	 	&	 W. Pietsch	 	&	 2010-01-25	 	&	 00:42:44.3	 	&	 41:16:09	 	&	 --	  	&	 --	  	&	 15460	 \\ 
0600660601	 	&	 M31	 	&	 W. Pietsch	 	&	 2010-02-02	 	&	 00:42:44.3	 	&	 41:16:09	 	&	 --	  	&	 15395	  	&	 --	 \\ 
0650560201	 	&	 M31	 	&	 W. Pietsch	 	&	 2010-12-26	 	&	 00:42:44.3	 	&	 41:16:09	 	&	 --	  	&	 20674	  	&	 22119	 \\ 
0650560301	 	&	 M31	 	&	 W. Pietsch	 	&	 2011-01-04	 	&	 00:42:44.3	 	&	 41:16:09	 	&	 19168	  	&	 28014	  	&	 28557	 \\ 
0650560501	 	&	 M31	 	&	 W. Pietsch	 	&	 2011-01-25	 	&	 00:42:44.3	 	&	 41:16:09	 	&	 --	  	&	 19174	  	&	 21824	 \\ 
0650560601	 	&	 M31	 	&	 W. Pietsch	 	&	 2011-02-03	 	&	 00:42:44.3	 	&	 41:16:09	 	&	 --	  	&	 21477	  	&	 21900	 \\ 
0672130101	 	&	 M32	 	&	 D. Wang	 	&	 2011-06-27	 	&	 00:42:41.8	 	&	 40:51:54	 	&	 53978	  	&	 78584	  	&	 84454	 \\ 
0672130501	 	&	 M32	 	&	 D. Wang	 	&	 2011-07-13	 	&	 00:42:41.8	 	&	 40:51:54	 	&	 15055	  	&	 24357	  	&	 28197	 \\ 
0672130601	 	&	 M32	 	&	 D. Wang	 	&	 2011-07-05	 	&	 00:42:41.8	 	&	 40:51:54	 	&	 57231	  	&	 70398	  	&	 72644	 \\ 
0674210201	 	&	 M31	 	&	 W. Pietsch	 	&	 2011-12-28	 	&	 00:42:44.3	 	&	 41:16:09	 	&	 17636	  	&	 19896	  	&	 19778	 \\ 
0674210401	 	&	 M31	 	&	 W. Pietsch	 	&	 2012-01-15	 	&	 00:42:44.3	 	&	 41:16:09	 	&	 15452	  	&	 18595	  	&	 18780	 \\ 
0674210501	 	&	 M31	 	&	 W. Pietsch	 	&	 2012-01-21	 	&	 00:42:44.3	 	&	 41:16:09	 	&	 --	  	&	 16535	  	&	 16603	 \\ 
0674210601	 	&	 M31	 	&	 W. Pietsch	 	&	 2012-01-31	 	&	 00:42:44.3	 	&	 41:16:09	 	&	 --	  	&	 15947	  	&	 16740	 \\ 
0690600401	 	&	 XBo 135	 	&	 R. Barnard	 	&	 2012-06-26	 	&	 00:42:52.0	 	&	 41:31:07	 	&	 53676	  	&	 84691	  	&	 90261	 \\ 
0700380601	 	&	 XMMU J004243.6+412	 	&	 N. Schartel (PS)	 	&	 2012-08-08	 	&	 00:42:43.7	 	&	 41:25:18	 	&	 17841	  	&	 18819	  	&	 19241	 \\ 
0701981201	 	&	 M31N 2013-01b	 	&	 N. Schartel (PS)	 	&	 2013-02-08	 	&	 00:44:02.1	 	&	 41:25:44	 	&	 16461	  	&	 19306	  	&	 19179	 \\ 
0763120101	 	&	 M31 Disk 1	 	&	 M. Sasaki	 	&	 2015-06-28	 	&	 00:44:22.0	 	&	 41:31:16	 	&	 75830	  	&	 91377	  	&	 91557	 \\ 
0763120201	 	&	 M31 Disk 1	 	&	 M. Sasaki	 	&	 2016-01-21	 	&	 00:44:22.0	 	&	 41:31:16	 	&	 40080	  	&	 50817	  	&	 52017	 \\ 
0763120301	 	&	 M31 Disk 2	 	&	 M. Sasaki	 	&	 2015-08-11	 	&	 00:44:49.4	 	&	 41:49:31	 	&	 94854	  	&	 97252	  	&	 97317	 \\ 
0763120401	 	&	 M31 Disk 2	 	&	 M. Sasaki	 	&	 2016-01-01	 	&	 00:44:54.4	 	&	 41:49:12	 	&	 45297	  	&	 66417	  	&	 68747	 \\ 
\hline
\end{longtable}
\end{small}

\section{Spectral fits}
\begin{figure*}
\begin{center}
\resizebox{6.3in}{!}{\includegraphics[trim= 0cm 0cm 0cm 0cm, clip=true, angle=0]{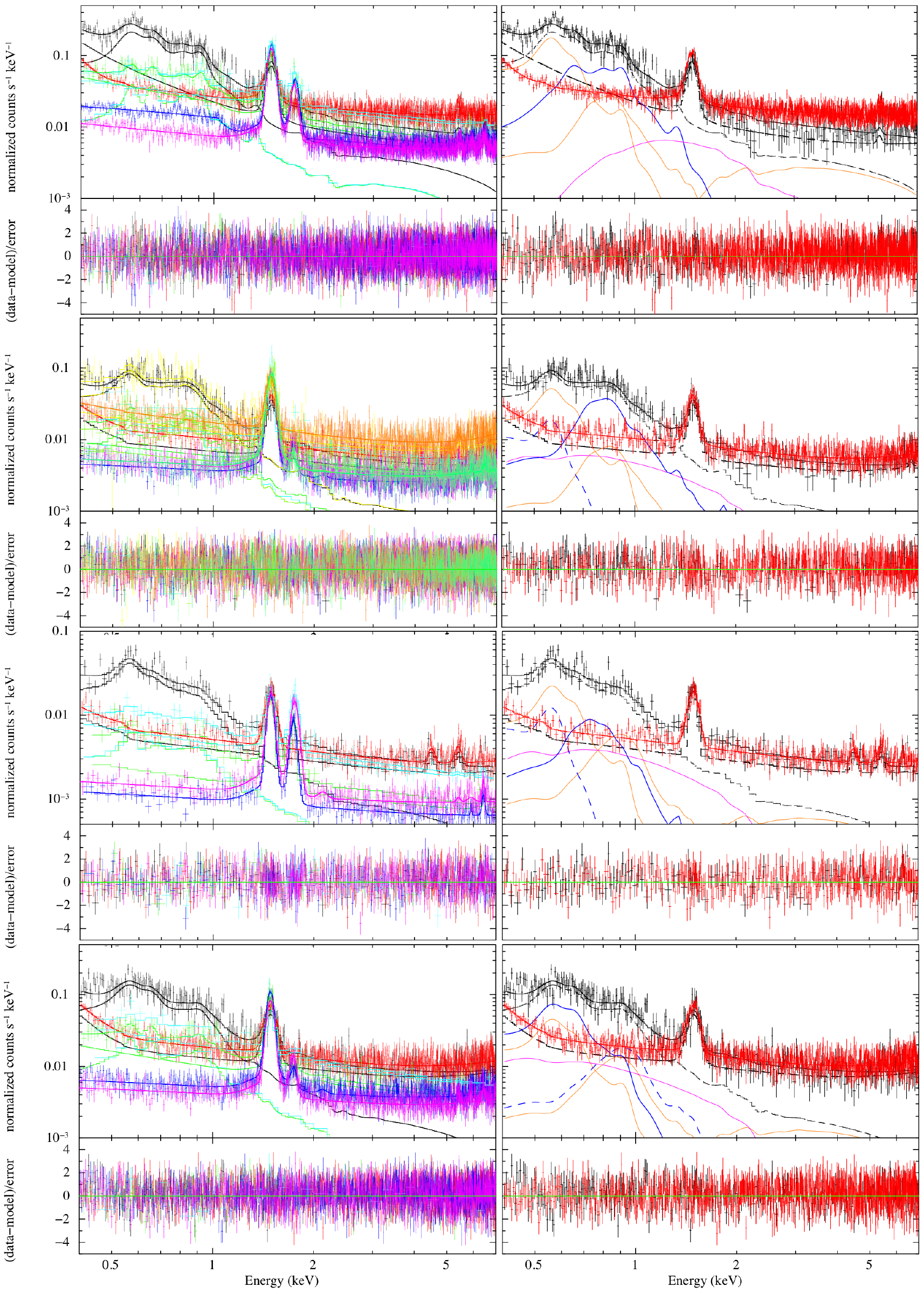}}
\caption{\xmm\ spectral fits of our extended emission Regions~1 through 4, arranged from top to bottom. The left panels show the fits to all spectra while the right panel shows the EPIC spectra with the most counts along with additive components of the best fit model for clarity. The red data points and line are the FWC data and best fit. The black data points and associated lines are the observational spectra, best fit model (solid), combined astrophysical components (dash-dot), and combined particle-induced components (FWC plus residual SP, dashed). The orange lines represent the AXB components, the magenta line shows the unresolved SNR, XRB, and SSS component (see Sect.~\ref{x-ray_background} for descriptions), blue solid lines represent the \gal\ ISM component with the dashed blue lines indicating the contamination of smaller regions by larger extended emission regions (e.g. Region~5 component in Region~2). Best fit parameters are given in Table~\ref{spectral_fit_results}.}
\label{spectral_fits1}
\end{center}
\end{figure*}

\begin{figure*}
\begin{center}
\resizebox{6.3in}{!}{\includegraphics[trim= 0cm 0cm 0cm 0cm, clip=true, angle=0]{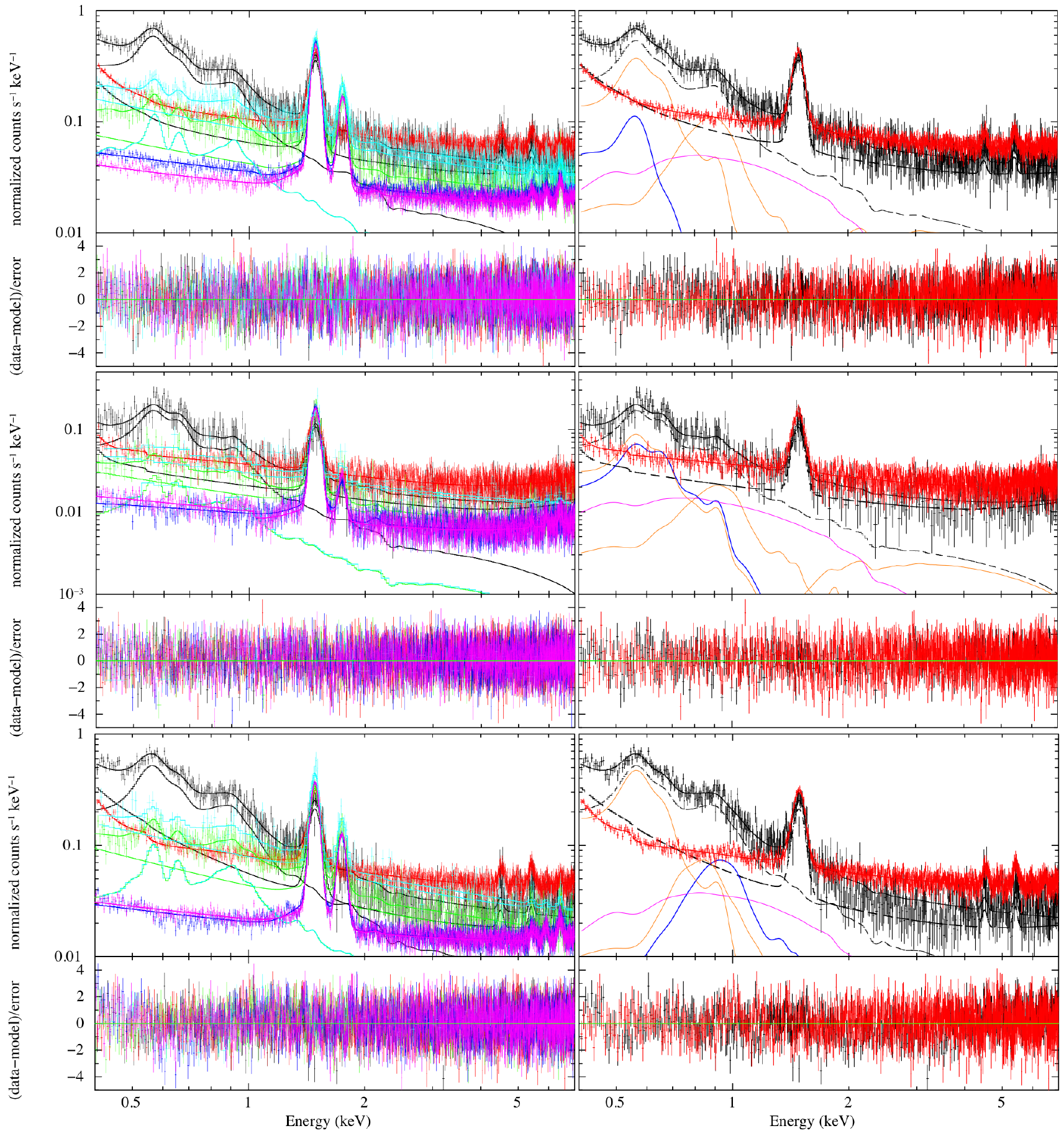}}
\caption{Same as Fig.~\ref{spectral_fits1} for Regions 5, 6, and 7, arranged from top to bottom. Best fit parameters are given in Table~\ref{spectral_fit_results}.}
\label{spectral_fits2}
\end{center}
\end{figure*}

\end{appendix}
\end{document}